\documentclass[sigconf, screen=true, authorversion]{acmart}

\usepackage{enumitem}

\setcopyright{acmlicensed}
\copyrightyear{2024}

\acmBooktitle{The 2024 ACM Conference on Fairness, Accountability, and Transparency (FAccT '24), June 3--6, 2024, Rio de Janeiro, Brazil}
\acmConference[FAccT '24]{The 2024 ACM Conference on Fairness, Accountability, and Transparency}{June 3--6, 2024}{Rio de Janeiro, Brazil}
\acmYear{2024}
\acmISBN{979-8-4007-0450-5/24/06}
\acmDOI{10.1145/3630106.3658957}

\begin{document}

\title{A Framework for Assurance Audits of Algorithmic Systems}

\def\bablai{
    \institution{BABL AI Inc.}
    \city{Iowa City} \state{Iowa} \country{USA}
}
\def\lmu{
    \institution{Ludwig Maximilians University}
    \city{Munich} \country{Germany}
}
\def\arva{
    \institution{AI Risk and Vulnerability Alliance}
    \city{Seattle} \state{Washington} \country{USA}
}
\def\uoi{
    \institution{University of Iowa}
    \city{Iowa City} \state{Iowa} \country{USA}
}

\author{Khoa Lam}
\email{khoalam@bablai.com}
\authornote{Authors contributed equally to this research.}
\orcid{0009-0004-8970-4537}
\affiliation{\bablai}

\author{Benjamin Lange}
\authornotemark[1]
\email{benjamin.lange@lmu.de}
\orcid{0000-0002-5809-8704}
\affiliation{\bablai}
\affiliation{\lmu}

\author{Borhane Blili-Hamelin}
\authornotemark[1]
\email{borhane@avidml.org}
\orcid{0000-0002-9573-3332}
\affiliation{\bablai}
\affiliation{\arva}

\author{Jovana Davidovic}
\email{jovana-davidovic@uiowa.edu}
\orcid{0000-0002-8998-5496}
\affiliation{\bablai}
\affiliation{\uoi}

\author{Shea Brown}
\email{sheabrown@bablai.com}
\orcid{0000-0002-6451-9675}
\affiliation{\bablai}
\affiliation{\uoi}

\author{Ali Hasan}
\email{ali-hasan@uiowa.edu}
\orcid{0000-0003-2963-2573}
\affiliation{\bablai}
\affiliation{\uoi}

\renewcommand{\shortauthors}{Lam, et al.}

\begin{abstract}

    An increasing number of regulations propose ‘AI audits’ as a mechanism for achieving transparency and accountability for artificial intelligence (AI) systems. Despite some converging norms around various forms of AI auditing, auditing for the purpose of compliance and assurance currently lacks agreed-upon practices, procedures, taxonomies, and standards. We propose the ‘criterion audit’ as an operationalizable compliance and assurance external audit framework. We model elements of this approach after financial auditing practices, and argue that AI audits should similarly provide assurance to their stakeholders about AI organizations' ability to govern their algorithms in ways that mitigate harms and uphold human values. We discuss the necessary conditions for the criterion audit and provide a procedural blueprint for performing an audit engagement in practice. We illustrate how this framework can be adapted to current regulations by deriving the criteria on which ‘bias audits’ can be performed for in-scope hiring algorithms, as required by the recently effective New York City Local Law 144 of 2021. We conclude by offering a critical discussion on the benefits, inherent limitations, and implementation challenges of applying practices of the more mature financial auditing industry to AI auditing where robust guardrails against quality assurance issues are only starting to emerge. Our discussion---informed by experiences in performing these audits in practice---highlights the critical role that an audit ecosystem plays in ensuring the effectiveness of audits.

\end{abstract}



\maketitle

\section{Introduction} \label{intro}
    
    Auditing has been proposed in a variety of laws and regulations \cite{nyc_ll144,amerprivact,stopdiscact,algoaccact,ftc,aiact,dsa},\footnote{\citet{mokander_22} argue that \textit{conformity assessments} as an enforcement mechanism in the EU AI Act should be interpreted as auditing, despite the term ‘audits’ not being referred to explicitly.} standardized frameworks \cite{nist_airmf, schwartz_22}, and guidelines for industry best practices \cite{baxter_21,hhs} as a mechanism to identify and mitigate risks of harm in artificial intelligence (AI)\footnote{In this paper, we use ‘AI,’ ‘algorithm,’ ‘AI system,’ and ‘algorithmic system’ interchangeably. Similarly, ‘AI audit’ and ‘algorithm audit’ both convey the same meaning.} and to build public trust and promote accountability for AI system developers and deployers. Most notably, New York City (NYC) enacted the Local Law 144 of 2021 (‘NYC bias audit law’ hereafter) in which \textit{bias audits}---defined as independently conducted impartial evaluations---are required for \textit{automated employment decision tools} used in hiring and promotion \cite{nyc_ll144}. Auditing for the purpose of compliance and assurance with normative requirements currently lacks defined norms and standardized practices, despite notable emerging efforts to perform audits as many types of engagement, including adversarial pressure testing \cite{angwin_16,snow_18,wall_21}, quantitative technical assessments \cite{wilson_21,engler_21,buolamwini_18,larson_16,chouldechova_18}, and qualitative assessments of risks or impacts \cite{brown_21,hasan_22,costanzachock_22}, among others.

    In this paper, we propose a criterion audit framework for the external compliance and assurance audits of algorithmic systems. This approach is inspired by how financial audits are used to provide assurance that financial statements are presented accurately and in conformity with generally accepted accounting standards (GAAS) \cite{guszcza_18}. Our aim is to show \textit{how} and---in part---\textit{to what extent} the methodology and practices of such a mature industry can be applied to AI auditing as an emerging industry absent of agreed-upon standards and official certification bodies. While the idea to follow the financial auditing practices---at least in part---is in itself not new \cite{mokander_23_ds,raji_22,fh_manual}, our paper offers an actionable approach for performing external compliance and assurance audits which is missing in the literature. Furthermore, as we have also used this framework to perform audits for the NYC bias audit law, our critical discussion draws on practitioner perspectives from operationalizing these external audits in the field.
    
    Our paper proceeds as follows. Section \ref{background} provides relevant background on the AI auditing landscape, including the operational gaps in the current regulations, and a short survey on taxonomy. We then introduce the criterion audit framework in detail. Section \ref{audit} outlines the (1) primary objectives of the audit framework, (2) its key elements, (3) the approaches to auditing supported by this framework, (4) the procedures for executing this audit, and lastly, (5) the responsibilities of an auditor tasked with its performance. In Section \ref{illustration}, we demonstrate a specific use case of the framework by adapting it to NYC Local Law 144 of 2021. Section \ref{benefits} critically examines the benefits, limitations, and implementation challenges posed by this framework. Finally, section \ref{conclusion} concludes our discussion.

\section{Background} \label{background}

\subsection{Motivation}

    In recent years, policymakers have introduced various forms of formal evaluation of algorithmic systems as a prominent mechanism to bring about transparency and accountability, with some requiring that these algorithmic systems undergo ‘audits’ \cite{nyc_ll144,amerprivact,stopdiscact,algoaccact,ftc,aiact,dsa}. However, most proposed policies have offered little guidance on audit quality assurance issues, such as independence rules, engagement performance matters, auditor qualification, and quality control procedures \cite{lucaj_23}. Among these, legislation has focused predominantly on independence. While some regulations, such as NYC bias audit law and EU Digital Services Act, have set rules around auditor eligibility, others, such as The Algorithmic Accountability Act of 2020 and the EU AI Act, offer no guidelines on independence at all.

    Standardized procedures remain largely unaddressed by regulations. Although there are emerging efforts to provide guidance for the execution of audits \cite{dsa_audit,aic4}, there is no consensus regarding how audits ought to be performed. Auditors are thus left to their own devices to operationalize audit engagements \cite{costanzachock_22}. This lack of clear guidance poses several challenges in the current algorithm auditing industry. First, there is a significant risk of inconsistency in audit quality, due to the discrepancies in audit engagement performance. This issue can eventually lead to failure to achieve the aims of the regulation for which audits were performed. Second, without standard practices, auditors themselves may face issues of liability for risks concerning false assurance \cite{goodman_23}.

    Against this landscape, our discussion aims to (1) provide a framework that answers central questions about how auditing for the purpose of compliance or assurance can be conceptualized, structured, and applied in practice, and (2) critically consider its key benefits and limitations.\footnote{See \cite[p. 294--296]{goodman_23} for key questions that the audit regime has to answer.}

\subsection{Taxonomy} \label{taxonomy}

    Given the flexible nature of audits in other contexts, the term ‘audit’ has been borrowed to describe many disparate forms of evaluation of algorithmic tools, products, and systems \cite{goodman_23,costanzachock_22}. Audit engagements can range from pressure-testing efforts by journalists and civil society without direct access to the audited system \cite{angwin_16,larson_16,snow_18,wall_21},\footnote{We note that while journalists do not typically refer to their works as ‘audits’ but as ‘analyses’ or ‘investigations,’ these efforts are nonetheless widely cited as exemplary audits in the reviewed literature.} to pre-deployment evaluations conducted by teams internal to an organization \cite{raji_20}, to ‘collaborative audits’ conducted by an outside team without safeguards against conflicts of interest \cite{wilson_21}, to audits conducted by outside parties with access to a system under robust safeguards against conflicts of interest \cite{raji_22}. Here, we focus specifically on ‘external audits’ aimed at providing assurance of compliance with requirements set forth in legislation or other standardized frameworks, which are sometimes also referred to as ‘compliance audits’ \cite{adalovelace,mokander_23}.

    Internal audits---such as the model evaluations required by the US Federal Reserve’s Supervisory Letter (SR) 11-7 \cite{sr117}---are forms of self-assessment: evaluation conducted by the audited party or by outside contractants without robust safeguards against conflicts of interest \cite{ajunwa_21,raji_22}. By contrast, external audits are evaluations conducted by independent parties outside of the audited entity, such as the public or regulators. In the US, relevant models of external audits from other fields include outside evaluations conducted by certified independent third parties---such as financial audits that conform to the Sarbanes--Oxley Act of 2002 \cite{sox} and Leadership in Energy and Environmental Design (LEED) certification by the US Green Building Council---and evaluations conducted by a government agency, such as FDA inspections for food and drugs \cite{ajunwa_21,raji_22}.\footnote{On the long history of independent audits in finance, see \cite{goodman_23,power_99,power_94}.} There is a growing consensus that, with internal audits, conflicts of interest between the auditor and the auditee may undermine their purpose of providing the assurance of trustworthiness or compliance to outside parties \cite{goodman_23,raji_22,ajunwa_21,sloane_21,dsa,nyc_ll144}. Potential measures for ensuring independence and preventing conflicts of interest in external audits include: rules against cross-selling of non-audit services, legal liability of auditors for false provision of assurance, standardization of audit criteria, professionalization of audits, creation of ‘auditing intermediaries’ \cite{wagner_21},\footnote{An ‘auditing intermediary’ refers to a proposed specialized audit entity whose responsibility is to ensure that accurate and verified data from large online platforms is received by regulators when shared across jurisdictions.} government selection of auditors, among others \cite{raji_22,goodman_23}.

    Evaluations performed by external parties without adequate access to the audited system---referred to as ‘critical third-party audits’ by \citet{metcalf_21}---can play an important role in uncovering and drawing public attention to existing flaws of deployed AI systems. They aim to inspire accountability by way of forcing organizations to ‘fix’ their AI systems (through technological or governance means) or face public scrutiny. However, we follow \citet[p. 558]{raji_22} in assuming that this form of third-party oversight without adequate access “[is] not conventionally considered audits per se.” By contrast, our proposed audits require formal auditor access to the audited system. Furthermore, a more fundamental distinction concerns its public role, where criterion audits, similar to financial audits, function to provide ‘comfort’ by way of assurance and certification.\footnote{See \cite[p. 126]{power_99} for discussion on public function of financial audits. In AI auditing, tensions also arise from disagreements about which role audits ought to play \cite{mokander_23_ds}.}

    Finally, we also consider audits distinct from risk or impact assessments \cite{sloane_23}. First, proposals for algorithmic impact or risk assessments typically focus on internal assessments \cite{raji_22,selbst_21}. Second, as we examine in this paper, effective external audits require a determination of an outcome (e.g., whether an audit passes or fails) such that it can enable stakeholders of the audit to act accordingly. By contrast, risk or impact assessments are better understood as having open-ended outputs, such as a prioritized and normatively justified list of risks or impacts \cite{selbst_21, hasan_22}.

\section{The Criterion Audit} \label{audit}

\subsection{Objectives}

    The primary objective of financial auditing is to provide assurance to stakeholders (e.g., investors, lenders, and regulators) regarding the reliability of an organization’s financial information for informed decision-making. Financial auditing assumes a vital role in upholding public trust in the financial system and ensuring economic stability by holding organizations accountable for their financial reporting and operations. External compliance and assurance algorithm audits should similarly aim to offer stakeholders reassurance that algorithmic systems are designed, built, deployed, and governed in a responsible and transparent manner. While AI systems do not bear foundational responsibility for upholding the integrity of the financial market, we argue that the opaque, rapidly transformative, and increasingly pervasive nature of AI systems in everyday life demands an assurance that algorithmic systems function in ways that mitigate harms and uphold human values.\footnote{There is convergence on the importance of external audits for providing assurance to outside parties, such as to governments, the public, or other organizations \cite{goodman_23,raji_22,ajunwa_21,sloane_21,mokander_23}.} AI audits can thus be employed as a mechanism to provide this desired reassurance and trust between stakeholders of AI by way of systematization and transparency \cite{mokander_23_ds}.

    If transparency is a key mechanism for AI audits to foster trust, current practices in the algorithm auditing industry often fall short of achieving such an ideal. Most auditing efforts do not publish outcomes or procedures to the public, often due to clients' confidentiality agreements \cite{costanzachock_22}. Many proposed regulations of AI auditing also do not require disclosure of an outcome. This lack of transparency can gradually degrade accountability by rendering auditing toothless, as audited organizations are not incentivized through either public pressure or regulatory enforcement to create meaningful change. Furthermore, due to procedural non-transparency, audit process effectiveness is not measurable, perpetuating what \citet{power_94} refers to as the ‘regress of mistrust,’ where accountability and trust are shifted away from the audited organizations and onto auditors themselves, who are in turn subjected to more audits and evaluations. Our proposed audit framework aims to provide this critical and meaningful layer of transparency by way of public disclosure requirements (see \ref{definition} and \ref{publication}). While public disclosure is not a comprehensive solution to achieve accountability,\footnote{On the limits of extensive public disclosure in financial audits, see \cite[p. 20]{power_94}.} our proposal aims to be sufficiently consequential in a way that it enables audit stakeholders (e.g., regulatory enforcers, audit report readers, impacted communities, the public, and developers/deployers themselves) to take actions towards advancing accountability for AI companies around the design, development, deployment, and governance of their algorithms.

\subsection{Definition} \label{definition}

    We define a \textit{‘criterion audit’} as:

    \begin{quote}
        
        \textbf{Criterion Audit:} A criteria-based independent external evaluation \textit{E} of an algorithmic system \textit{S} conducted by an auditor \textit{A} to determine whether the given system \textit{S} meets the requirements set by a normative framework.
    
    \end{quote}
    
    Audit criteria are the set of verifiable or observable conditions that must be jointly satisfied for an algorithmic system \textit{S} to count as compliant with a given standard or law. These conditions must enable auditors to form an unambiguous opinion about whether a given criterion is satisfied based on evidence they can obtain within the operational context of the audit. For example, as we show in later sections, in the case of NYC bias audit law, these may include the relevant criteria for determining how a given system is assessed for disparate impact. By ‘independent third-party evaluator,’ we refer to an auditor that is independent of the customer-supplier relationship and is free of any conflict of interest with respect to the client that is being audited. On this matter, we understand ‘independence’ as either: (1) lacking any contractual relationship with the auditee, or (2) involving rigorous safeguards against conflicts of interest, in cases where fees related to the audits themselves are paid by the auditee.\footnote{Such is the norm in financial audits, see also \cite{acc_handbook,acc_sec} for standards on independence for financial auditors.}

    Four features of the criterion audit are worth highlighting:

    \begin{enumerate}
    
        \item \textbf{The audit must be conducted against a set of standardized and publicly accessible criteria}, to provide procedural transparency to the audit. By showing what criteria the algorithmic system was evaluated against, the results of the audit are thus contextualized and more interpretable by readers of the audit report.\footnote{For non-financial assurance engagements, the International Standards on Assurance Engagements (ISAE) 3000 \cite{isae3000} specifies the characteristics of suitable audit criteria, which include relevance, completeness, reliability, neutrality, and understandability.}
        
        \item \textbf{The audit's objective must be to measure compliance or to provide assurance against a normative framework}, such as a regulation. This precise and narrow audit scope would allow for measurement and comparison of the audit’s performance \cite{raji_22,costanzachock_22,mokander_23_ds}. Moreover, it also provides the basis for which audit criteria are constructed.

        \item \textbf{Auditors must be trained and accredited in a standardized manner} \cite{raji_22,costanzachock_22}. They must also be held to high professional standards of quality assurance and ethics---e.g., in the form of a Code of Ethics or a standard of audit quality assurance. While audit standards are nascent, training courses and programs have started to appear \cite{certaied,fh_uni,bablcert}, aiming to provide such standardized certification for auditors.

        \item \textbf{Results of the audit must be publicly disclosed, at least in some restricted form}, to meet both the public’s need for standard reporting and audited organizations’ concerns regarding security and intellectual property \cite{raji_22}. Furthermore, the content of public disclosure can be stipulated by the legislation and supplemented further to facilitate contextual interpretation of the results. The degree of public disclosure can vary, depending on contextual details. However, at a minimum, it should contain key information that is necessary for readers of the audit report to understand compliance with a given piece of legislation.
        
    \end{enumerate}

\subsection{Auditing approaches} \label{approaches}

    A number of auditing approaches can be adapted to be compatible with the features of this audit framework. We introduce two paradigm approaches below:\footnote{These approaches are based loosely on ‘direct engagements’ and ‘attestation engagements’ in financial audits.}

    \subsubsection{Direct assessment.} The auditor directly performs the procedures on the algorithmic systems, as scoped by the audit. In this approach, the auditor is given---for technical testing---full, or mediated, access to the model to conduct technical testing, or---for a risk assessment---access to relevant stakeholders of the model, such as its impacted users. When the assessments are performed as part of a regulation's requirements, the criterion audit framework also demands (1) that these assessments be performed in accordance with a transparent set of criteria by qualified audit professionals, and (2) that their results be partially disclosed publicly.

    \subsubsection{Indirect verification.} Alternatively, the auditee can perform the assessments themselves, and subsequently submit a body of evidence to prove the performance and quality of these auditable procedures. This requires the auditors to evaluate the procedures, policies, and institutional structures performed and established by the auditee against a set of criteria that examines their quality. This process can act as a complement to the work of an internal audit team that evaluates the algorithmic system along its development lifecycle,\footnote{See also \cite{schuett_22} for discussion on The Three Lines of Defense (3LoD) model as a proposed mechanism for risk mitigation in the AI context.} and produces critical documents and artifacts---such as model or ethical risk analysis results, system cards, and technical testing reports \cite{raji_20}. In this way, efficiency is afforded between internal and external auditing functions where the latter reviews and evaluates documentation provided by the former for the specific objective of demonstrating compliance or assurance. Moreover, the criterion audit framework can apply if (1) the evaluation process is conducted by qualified practitioners, and (2) some results of the audit are subsequently publicly disclosed.
    
    There are inherent and practical trade-offs between these approaches. Direct assessment may provide better safeguards against accountability concerns such as rubber stamping, whereas indirect verification may find efficiency in large-scale audit engagements for auditees with internal audit teams. In practice, auditors may utilize one single approach, leverage both in a hybrid manner, or determine the suitable auditing approach based on an analysis prior to performing audit procedures.\footnote{In the Delegated Regulation on auditing for the Digital Services Act \cite{dsa}, auditors are required to conduct an audit risk analysis to select the precise audit methodologies.}
    
    In this paper, we propose procedures (in Section \ref{procedures}), and auditor responsibilities (in Section \ref{auditor}) based on the indirect verification approach. In addition, the criteria designed for NYC bias audit law (in Section \ref{criteria}) were also developed assuming the verification approach of auditing. Beyond the efficiency benefits mentioned above, we also see potential positive implications for audit independence. Auditors who themselves perform direct assessments may arguably lack the critical distance to provide an impartial and credibly independent evaluation of the rigor of their own testing. Indirect audits can thus help to mitigate such risk by introducing a separation between the party responsible for conducting the assessments and the party independently providing the assurance of their quality.

\subsection{Audit procedures} \label{procedures}

    We propose the following standard process to perform a criterion audit:

    \subsubsection{Target scoping.} The auditor scopes the audited algorithmic system to obtain a foundational understanding of its technical and sociotechnical components, in addition to any process deemed relevant to the normative framework, such as whether the system has undergone bias testing. This is intended to provide auditors with a contextual knowledge of the algorithm before evaluating it against a set of criteria. Scoping can be done, for example, in the form of a targeted questionnaire or through interviews with personnel of the audited organization.
    
    \subsubsection{Documentation submission.} The auditee submits documentary evidence towards satisfaction of the audit criteria, and the auditor reviews the documentation and determines whether the criteria have been preliminarily met, pending verification of evidence (see below). During the review process, if the auditor requires more evidence, they may ask for additional documentation or request interviews with the auditee’s internal or external stakeholders, such as employees or third parties, to facilitate evaluation. The auditor reaches an initial opinion about whether the criteria have been satisfied by the evidence provided by the auditee. Auditors, therefore, need to have the knowledge and expertise to make qualified critical judgments based on evidence of whether a given criterion has been sufficiently met.
    
    \subsubsection{Evidence verification.} The auditor determines the truthfulness of the evidence presented by the auditee. Methods for verification, such as examining official communication logs, or observing the re-performance of a computation in an interview, can vary based on the nature and importance of the evidence. For example, the auditee may show the computation of bias testing to verify the results shown as evidence submitted previously. At the end, the auditors reach a final opinion about the satisfaction of the criteria, which takes into account the veracity of the evidence.\footnote{In practice, some information required to verify evidence may have already been provided as part of the documentation submission (e.g., communication logs, database queries, event recordings), in which case, the auditor can now determine whether the quality of such evidence is sufficient.} This mirrors the standard procedures for obtaining evidence in financial audits \cite{as1105} and builds upon proposals calling for adaptation of claim-based assurance frameworks to the AI context \cite{hauer_23,brundage_2020}. As auditors also make judgment calls, they need to adhere to strict and rigorous standards in performing the verification.
    
    \subsubsection{Publication of the audit report.} \label{publication} The auditor drafts and publishes the audit report regarding the audited algorithmic system. Results should be disclosed in a standardized format, and the report at the least needs to explicitly and publicly show (1) whether each criterion has been met, (2) the final outcome of the audit, and (3) a formal opinion of the auditor. An audit report should also contain other standard information, which can include: in-scope and out-of-scope, a description of the algorithm,\footnote{This description should be at an appropriate level of specificity to facilitate readers' understanding of the audit quantitative results.} auditor's and auditee's general responsibilities, a statement on auditor independence, the level of assurance for the engagement (e.g., limited or reasonable assurance), and an informative summary of the work performed.\footnote{See \cite[p. 21, 60--65]{isae3000} for content of a standard financial audit report. See also \cite{8f_23,ripple_23} for our publicly available audit reports for NYC bias audit law.} Before publication, the auditee may be allowed to review the full report for strictly factual errors or any omissions that may have materially affected the report.
    
    \subsubsection{Certification.} The audited algorithmic system receives a certification indicating the outcome of the audit---e.g., whether the algorithmic system has passed (or failed) with respect to the targeted regulation against which it was evaluated.

\subsection{Auditor responsibility} \label{auditor}

    In financial auditing, the outcome of an audit requires an auditor to form an opinion on the financial statements based on having obtained sufficient appropriate audit evidence about whether these statements are free from material misstatement \cite{audit_q}. Achieving a high-quality audit requires auditors to (1) exhibit professional values and ethics, (2) have sufficient knowledge and skills in their subject matter, and (3) apply rigorous and appropriate audit processes and quality control procedures.
    
    An auditor for the criterion audit similarly bears the responsibility of forming an opinion based on the audit evidence provided by the auditee to evaluate the level of appropriateness, sufficiency, and material misstatement for the submitted body of evidence.\footnote{The Delegated Regulation on independent auditing under the Digital Services Act \cite{dsa_audit} identifies three specific audit risks: (1) inherent risks, (2) control risks, and (3) detection risks.} Determination of appropriateness, sufficiency, and materiality can take many forms, such as conducting statistical testing (e.g., statistical significance, power analysis), evaluating the appropriateness and reasonableness of a normative justification for important decisions (see \ref{diq}), or determining whether an auditee attempts to deceive the auditors or game the audit process. Such judgment calls and decision-making should therefore be made by auditors having not only (1) expertise in specific subject matter (such as technical acumen for technical audits, or expertise in normative ethics or sociology for sociotechnical audits), but also (2) training and certification in standardized audit process and quality control procedures, similarly to the audit quality training required for financial auditors.
    
    Research on algorithm auditing has so far called for auditor professionalization and certification as one of the requirements for high-quality audits \cite{raji_22,costanzachock_22}. However, there is little discussion on \textit{which} knowledge or training algorithm auditors should possess to perform audits effectively. Here, we advocate specifically for standard audit processes and quality control procedures as a fundamental requirement for auditors performing criterion audits. In practice, this form of training and industry knowledge can leverage the vast existing body of work from the financial auditing industry, where standardized methods have been developed to support auditors in audit engagements---such as Auditing Standard (AS) 1105 on audit evidence \cite{as1105}, the International Standard on Auditing (ISA) 315 on risks of material misstatement \cite{isa315}, and the International Auditing and Assurance Standards Board's standard on quality management for auditing firms \cite{isqm}). Furthermore, this training should also be complemented by comprehensive responsible AI education, such as on rigorous technical testing and effective risk management practices, to equip auditors with substantive domain knowledge when navigating complex considerations.

\section{Illustration: Adapting the Criterion Audit for NYC Bias Audit Law} \label{illustration}

    In November 2021, the New York City Council passed Local Law 144 of 2021 which requires bias audits for automated employment decision tools (‘AEDTs’) used in hiring and promotion.\footnote{\textit{Automated employment decision tools} are defined as “any computational process, derived from machine learning, statistical modeling, data analytics, or artificial intelligence, that issues simplified output, including a score, classification, or recommendation, that is used to substantially assist or replace discretionary decision making for making employment decisions that impact natural persons”.} The final rule, effective in July 2023, defines \textit{bias audit} as “an impartial evaluation by an independent auditor,” and the audit requires, as the minimum, an assessment of the tool’s disparate impact on persons of any component 1 category---i.e., race/ethnicity and gender categories. Independence rules are also established, in which auditors are not allowed to have been involved in using, developing, and distributing the AEDT, or to have contractual and financial interest in the organizations using, developing, and distributing the AEDT.

\subsection{Audit criteria development} \label{criteria}

    Applying the criterion audit framework, we derived a set of criteria which aims to determine whether an algorithmic system has met the requirements of NYC bias audit law. The audit criteria were developed as a function of (1) the legislation's content and specifications, and (2) our practitioner experience in directly conducting technical bias testing and ethical risk and impact assessments on algorithmic systems. Moreover, the audit criteria are constructed using the previously discussed indirect verification approach (see Section \ref{approaches}).
    
    The law requires ‘bias audits,’ at a minimum, to include a disparate impact analysis of the automated tool. Such technical analysis being the only explicit requirement for an audit stands in stark contrast to the broadly adopted view of current scholarship that mathematical-only perspectives of bias are insufficient in capturing impacts of algorithmic bias or in preventing discriminatory outcomes \cite{schwartz_22}. Moreover, the reviewed literature has documented extensively the limitations of a technical-only perspective in algorithmic bias and discrimination \cite{kim_17,watkins_22}, the importance of sociotechnical views of bias \cite{barocas_16}, the interdependence between technical and sociotechnical views in auditing of algorithmic systems \cite{hasan_22}, and the limitations of operationalizing technical measures in non-AI contexts \cite{danielsson_02,biagioli_16}.
    
    Adopting these sociotechnical views on bias management and mitigation, we designed our audit criteria to include three sections: (1) disparate impact analysis, (2) governance, and (3) risk assessment. For disparate impact analysis, we derived, to the best of our knowledge and ability, the minimally sufficient criteria allowing an auditor to evaluate a good faith analysis of the tool’s disparate impact performed by the audited organization. For governance, a set of requirements aims at evaluating the existing governance body within the auditee’s organization who is responsible and accountable for the management of risks related to bias of the AEDT. For risk assessment, we derived minimal requirements for an assessment of risks conducted by the auditee for the tool with a specific focus on bias.

    We view the governance and risk assessment requirements as prerequisites for a sufficiently rigorous disparate impact analysis prescribed by the law. Providing assurance for the rigor of this analysis---i.e., whether it runs the risk of being unreliable, being ill-informed, or lacking due diligence---requires investigating also the context within which the analysis was performed \cite{hasan_22,nist_airmf,selbst_19}. These criteria thereby serve as the bare minimum contextual factors that auditors should account for, namely (1) the organization's approach to \textit{controlling} disparate impact risks (governance), and (2) the organization's approach to making decisions about harm and bias mitigation priorities (risk assessment).

\subsection{Sections of an audit}

    Table \ref{tab:criteria} shows our set of audit criteria which aims to evaluate an algorithmic system for compliance with NYC bias audit law. The full set of criteria including all sub-criteria is available in Appendix \ref{full-criteria}. These criteria provide the basis for both reviewing submitted documentation and verifying evidence (see \ref{procedures}). While the execution of this audit can be tackled by multiple auditors (e.g., based on their areas of domain expertise), the three audit sections are not intended to be independent but rather complementary, both conceptually and operationally, to each other. This holistic approach is designed to consider technical bias risks in the sociotechnical context of the algorithm and governance measures.

    \begin{table*}[ht]
        
        \caption{Audit criteria for NYC Local Law 144 of 2021.}
        
        \label{tab:criteria}
        \begin{tabular}{lp{0.8\textwidth}}
            \toprule
            ID & Criterion \\
            \midrule
            \textbf{Q} & \multicolumn{1}{c}{\textbf{Disparate Impact Analysis}} \\
            \midrule
            Q.A & The tool analyzed for disparate impact shall be defined. \\
            Q.B & The dataset based on which disparate impact is analyzed shall be defined and characterized. \\
            Q.C & The demographic categories for which disparate impact can be analyzed using the dataset shall be defined. \\
            Q.D & Where the selection rate method is used, positive and negative outcomes of the tool shall be clearly defined as the basis for selection rate. \\
            Q.E & A metric which corresponds to selection rate or scoring rate shall be defined. \\
            Q.F & The ‘favored group’ and ‘disfavored groups’ \cite{ugesp} shall be identified, for all demographic categories. \\
            Q.G & The impact ratios shall be disclosed for all disfavored groups, for all demographic categories. \\
            Q.H & Where the selection rate method was used, statistical significance calculations of the difference between selection rates shall satisfy Uniform Guidelines on Employee Selection Procedures (UGESP) \cite{ugesp}. \\
            \midrule
            \textbf{G} & \multicolumn{1}{c}{\textbf{Governance}} \\
            \midrule
            G.A & The auditee shall have a party which is accountable for risks related to disparate impact. \\
            G.B & The duties of the party accountable for risks related to disparate impact shall be clearly defined. \\
            G.C & The auditee shall provide evidence that the defined duties of the party accountable for risks related to disparate impact are carried out. \\
            \midrule
            \textbf{R} & \multicolumn{1}{c}{\textbf{Risk Assessment}} \\
            \midrule
            R.A & The auditee shall have completed a risk assessment of the tool. \\
            R.B & The risk assessment shall show identification of relevant risks related to bias. \\
            R.C & The risk assessment shall demonstrate appropriate evaluation of relevant risks. \\
            \bottomrule
        \end{tabular}
    
    \end{table*}

    \subsubsection{Disparate impact analysis.} \label{diq} The law requires the disclosure of the disparate impact assessment results in the form of impact ratios for groups of race/ethnicity, sex, and their intersections. An auditee has to have made a number of key decision points to arrive at the quantitative results. These subjective and discretionary decision points mirror concepts known in the qualitative research as ‘choice moments’ \cite{savinbaden_12}. Our criteria was accordingly designed to elucidate these decision points from the auditee and provide auditors a mechanism to evaluate their appropriateness.
    
    The auditors evaluate: (1) the definition of AEDT by the auditee (Q.A), (2) the dataset used for analysis (Q.B), (3) the demographic information and its data collection process (Q.C), (4) the selection or scoring rate definition and its basis (Q.D \& Q.E), (5) demographic groups covered in the analysis (Q.F), (6) impact ratio calculations, including of uncertainties (Q.G), and (7) statistical significance calculations of the difference between selection rates (Q.H).
    
    For each criterion, a set of sub-criteria provides further guidance for auditors to assess the quality of the auditee decisions (see Appendix \ref{full-criteria}). For instance, sub-criteria Q.D.1 requires the auditee to provide the justification for their choice of positive outcome as the basis for selection rate.\footnote{‘Positive outcome’ refers to the favorable outcome for a candidate from the use of the model, such as being selected to move forward in the hiring process or assigned a classification by an model.} An auditor---having subject matter competency in a technical domain such as data science in combination with experience in algorithmic bias work---would be able to then evaluate whether, for example, using interview rates or hiring rates resulting from the AEDT use is more appropriate for calculations.
    
    \subsubsection{Governance.} Although the auditing of governance structures of an AEDT is not specified as requirements by the law, this set of criteria requires that the auditee has an accountable party for risks related to bias in a way that is clearly defined and operationalized across the organization.
    
    Regulatory guidance and industry best practices have emphasized the role of governance and internal controls as a foundational building block of AI risk management systems \cite{nist_airmf,gao,occ}. Furthermore, our practitioner experience in performing audit-type engagements has also led us to believe that effective mitigation of the risks posed by an organization's use, development, and deployment of algorithm (including ones related to bias) requires some designated oversight body within the organization that is accountable for them. The rationale for including these criteria was therefore to determine whether this minimally sufficient layer of accountability is established.
    
    The criteria require: (1) that the auditee has an accountable party for risks related to disparate impact (G.A), (2) that the responsibilities of this party are clearly defined (G.B), and (3) that such responsibilities have been carried out prior to the audit (G.C).
    
    Similar to those in disparate impact analysis, specifications for the evidence in the form of sub-criteria require the auditee to show formalization of such governance. For instance, sub-criteria G.B.2 requires the auditee to show that the accountable party must have influence (e.g., through institutional power) over product changes. Evidence for this section can take the form of policies and procedures related to internal oversight---such as the charter of a responsible AI team or AI risk committee, their duties, and testimony by designated parties in the organization.
    
    \subsubsection{Risk assessment.} Scholarship has highlighted that technical measures of risks of harm (such as bias) need to be understood in the sociotechnical context of an algorithmic system \cite{sloane_23,nist_airmf,adalovelace,selbst_19}. This set of criteria examines the degree to which awareness for these sociotechnical risks is shown by the auditee.
    
    Specifically, criteria in this section require (1) that the auditee has completed a risk assessment of their AEDT (R.A), (2) that risks have been identified (R.B), and evaluated (R.C), as part of this assessment. To satisfy these criteria, the auditee is required to show, in their risk assessment, sufficient awareness of sociotechnical risks of harm, and of bias in particular. Moreover, auditors tasked with evaluating these criteria can further verify the quality of this risk assessment in a verification interview where contributors of the risk assessment may be invited to speak on the details of the risks and their justifications.

\subsection{Interactivity between sections of the audit}

    The three sections (i.e., disparate impact analysis, governance, and risk assessment) are intended to complement each other, reflecting the interwoven nature of the various aspects of AI risk management. \citet{hasan_22} provides an exemplary case for how this interplay between these areas should play out on the side of the audited entity when performing these auditable procedures. By identifying and evaluating risks related to bias, the auditee gains a sociotechnical understanding of their AEDT, which then informs whether and which risks can be measured using the available data, and how this technical testing should be performed. This dependency should, as a result, also be reflected in the disparate impact analysis, whose metrics and heuristics are guided by the risk assessment, and whose results reciprocally inform risk mitigation measures.\footnote{Requirements of the EU AI Act for high-risk AI systems also reflect the importance of this interaction between technical testing and risk management, in which system providers are required to identify risk management measures through diligent system testing, disclose such testing heuristics, and justify their testing metrics.}
    
    We expect auditor evaluations of these components to also have this sense of interdependence. Consider the scenario where the auditee has identified, in their risk assessment, that a risk of bias is created by the way an automated system’s user interface used in hiring is paginated such that candidates appearing on the first page is significantly more likely to receive an interview or be contacted (e.g., by recruiters). The auditor evaluating the risk assessment is encouraged, upon discovery of this information, to inform the auditor tasked with the disparate impact analysis of this information, so that they may take into account this detail when evaluating the basis of the metric used to calculate impact ratios. The disparate impact auditor should accordingly assess whether this element is taken into consideration in the technical analysis at all, and whether their metric is appropriate given the identified risks. Moreover, this procedural feature operationalizes the integration between what \citet{mokander_23_ds} refers to as ‘narrow’ and ‘broad’ auditing, where ‘narrow’ is exemplified by technology-focused testing or assessments and ‘broad’ is characterized by process-focused review of management of the technology.

\section{Benefits \& Limitations} \label{benefits}

\subsection{Benefits}

    There are three key benefits to the proposed criterion audit:

    \subsubsection{Adaptability.} Target scoping enables the audit to be tailored to a variety of algorithm use cases and legislative works. By stipulating minimally necessary and jointly sufficient conditions, the audit is a resource-efficient mechanism that can have a high degree of impact to provide assurance for a specific algorithmic system.
        
    \subsubsection{Efficiency and scalability.} The proposed framework provides an efficient and scalable method for auditing. In many cases, organizations rely on dozens to hundreds of algorithms, which presents a particular challenge for effective governance and assessments. By performing AI audits in a systematic and transparent manner, our proposed approach ensures that a larger number of algorithms can be audited in a consistent and clearly defined set of procedures.
        
    \subsubsection{Transparency and accountability.} Public disclosure of audit results and audit criteria provides a high degree of transparency. Considering the nascent stage of algorithm auditing as an industry, such level of transparency can be a powerful mechanism to foster accountability by (1) making the effectiveness of our proposed methodology measurable and comparable against other compliance and assurance frameworks, and (2) providing grounds for public scrutiny of AI companies concerning, among others, their adherence to regulations and ability to fulfill obligations to stakeholders and society.
        
\subsection{Implementation challenges} \label{implementation}

    There are a number of challenges in implementing the proposed audit framework:
    
    \subsubsection{Auditing standards development.} Developing a set of standards precisely scoped to a legislative work is not a straightforward or unambiguous task. As stated in Section \ref{approaches}, audit criteria can be built using various approaches to auditing. In addition, in cases where substantive differences of opinion exist between experts---such as in the discourse on computational definitions of fairness \cite{kleinberg_17,binns_20,bell_23}---designers of audit standards must navigate these nuances while also balancing to achieve a set of criteria that provides compliance and is operationalizable at scale.
        
    Another issue focuses on \textit{who} should be developing auditing standards. Standards are a powerful means to establish auditing norms and practices, and there are warranted concerns about corporate capture to shape them in accordance with their interests \cite{young_22}. In financial auditing, the International Financial Reporting Standards (IFRS) are developed and maintained by the independent International Auditing and Assurance Standards Board (IAASB). By contrast, auditing criteria for AI regulation compliance are currently emerging, where sets of criteria, despite being developed by independent standard-setting bodies \cite{fh_nyc,ieee,iso42001,aic4}, have yet gained wide adoption from the industry.

    \subsubsection{Training and gray area decision-making.} Following the training and accreditation of financial auditors, emphasis must be placed on equipping auditors with the competency and capabilities to perform these audits. As critical judgment is a fundamental feature of this methodology, auditors need to be equipped to better understand whether and, if so, the extent to which a given criteria may or may not be sufficiently met, or if further consultation is required. However, there is currently little discourse in the algorithm auditing field and limited training resources for aspiring algorithm auditors on this matter. The current state of auditor training unfortunately relies largely on on-the-job training, and self-empowerment for making judgment calls.
        
    \subsubsection{Opinion shopping and its effect on auditor independence.} As the AI auditing industry is in its infancy, there lacks a set of safeguards sufficiently robust to ensure high quality audits. Currently, there is little to none preventing auditors from making their audits as easy to ‘pass’ as possible to capture the market of organizations in scope for audits. This lack of safeguards enables audited entities to search for auditors who are more likely to provide a favorable opinion or whose audits are less stringent, known as \textit{opinion shopping} in the financial sector \cite{lennox_00,mccann_19,choi_18}. This practice creates an environment in which self-imposed substantial independence requirements for algorithm auditors become a business risk for auditing firms. In the case of NYC bias audit law, auditors are thus incentivized to maintain only the minimum independence requirements specified by the regulation, \textit{irrespective} of whether such specifications are sufficiently robust against deteriorating audit quality. This superficial sense of rigor can greatly perpetuate ‘audit-washing,’ whereby harms supposedly prevented by the audit are instead distracted from or even excused \cite{goodman_23}. This problem is in contrast to that in the financial sector, in which strong independence rule has two hundred years of history, precedent, and significant support from various stakeholders in the auditing ecosystem, including regulatory authorities.
        
\subsection{Limitations} \label{limitations}

    The criterion audit framework has a number of intrinsic limitations:
    
    \subsubsection{Reliance on the regulatory requirements.} The criterion audit is by design dependent on the formal requirements of a regulation or other standardized frameworks. Effective rulemaking and precise scoping of the requirements allow room for the audit and its criteria to become instrumental in meaningfully preventing or mitigating algorithmic harms. On the one hand, excessively broad scope risks the criterion audit not having enough teeth to enforce meaningful changes to the status quo. On the other hand, if the legal requirements are excessively restrictive, the criteria risk becoming a checklisting exercise for auditors, failing to achieve the objectives set forth by the law.\footnote{Policymakers face a similar challenge for the level of precision in rulemaking: if regulations are too specific, they cannot be appropriate for or adaptable to every given system; on the other hand, if too broad or general, they risk too easily passing the requirements.} For NYC bias audit law, the rule specifies various details on disclosure of impact ratios as a metric to assess disparate impact but does not require actions to address, manage, and mitigate bias and discrimination harms resulting from such tools in a systematic manner. This misdirected focus restricts the scope of the criteria greatly and thereby limits its ability to fulfill its intended goals.\footnote{NYC bias audit law relies entirely on employers to self-disclose a “hyperlink to a website containing the required summary of results” \cite{nyc_chapT}. See \cite{wright_24,groves_24} for an empirical investigation and discussions on the potential limitations of employer discretion within the law on achieving transparency and accountability goals.} Here, we urge regulators to pay close attention to the practitioner experience of auditors to understand to what extent the goals of their proposed rules can be met within the bounds of auditing mechanisms such as the criterion audit.
        
    \subsubsection{Reliance on an audit oversight ecosystem.} Modeled after financial auditing, this audit framework similarly requires a network of entities working in tandem to ensure its effectiveness. This network includes various stakeholders, from auditing firms, audited entities, entities developing standards, to entities providing training and certifying practitioners. Overseeing this network requires not only lawmakers, but also regulatory bodies who have the authority to appoint independent parties, certify standards, and enforce the law. As this ecosystem is currently not fully developed, audits performed using this audit framework may yet suffer from audit quality issues.
        
    \subsubsection{Reliance on a clearly defined outcome taxonomy.} In financial auditing, an audit outcome is expressed as an opinion by the auditor indicating whether the financial statements are presented fairly. There are four types of audit opinion; each one has a clear definition about the nature of the financial statements, and specifies what users should expect from the auditor in the audit report. A ‘disclaimer of opinion,’ for example, requires an auditor to explain why an opinion is withheld and to explicitly indicate that no opinion is expressed. Based on this information, the users of the audit report (e.g., shareholders) are thereby enabled to make appropriate financial decisions.
        
    This audit framework requires a similar set of outcome taxonomies, but there is no established or widely adopted equivalent in the AI auditing industry, at least to the extent that it meaningfully enables readers of the audit report to take appropriate actions. More concerning is that some regulations do not specify the disclosure of any audit outcome at all.\footnote{The only exception is the Digital Services Act, in which the Delegated Regulation on independent auditing \cite{dsa_audit} specifies the audit outcomes as either ‘positive,’ ‘positive with comments,’ or ‘negative,’ corresponding to the auditee’s extent of compliance.} Such is the case for NYC bias audit law, which does not require the automated tools to even \textit{pass} the audit to achieve compliance.\footnote{Compliance can simply be met when a tool, prior to its use: (1) has a bias audit conducted within one year, and (2) has a result summary of the most recent audit be made publicly available.} Moreover, without established industry norms around how audits ought to be used to make informed decisions, even attempts to implement a simple pass/fail opinion on audits may run the risk of diluting the significance of such judgment and rendering it difficult to interpret for audit stakeholders.

\section{Conclusion} \label{conclusion}

    In this paper, we have proposed the criterion audit framework for external compliance and assurance audits of AI systems. In the absence of standard practices for audits as proposed by emerging regulations, our approach offers an operationalizable auditing methodology aiming to provide assurance to audit stakeholders about organizations' ability to fulfill obligations about their algorithms. We lay out the necessary conditions for an effective criterion audit: an evaluation of an algorithmic system (1) against transparent audit criteria, (2) performed by qualified professionals, (3) with partially disclosed results to the public (4) about whether it complies with the requirements of a legislative work or standard. We offer procedures to perform the engagement for auditors, and highlight their responsibilities as evaluators of evidence about AI systems and the need for specialized training. We adapt the framework to NYC bias audit law, whereby we show the audit criteria to perform the audit, and provide the rationale and illustrative examples on auditor evaluation for each audit section, and the interplay between auditors tasked with different audit sections. We conclude by examining the benefits, implementation challenges, and inherent limitations of this audit framework, in part drawing from our perspective as auditors conducting bias audits under the NYC bias audit law. Our paper provides a glimpse into how auditing for compliance and assurance can fall short in a nascent industry even when modeled after practices of a more mature one. Our discussion emphasizes the critical need for an ecosystem---comprising auditors, auditees, regulators, standard-setting bodies, certification bodies, and enforcing agencies---working in tandem to ensure high-quality compliance and assurance external audits. Finally, even when such an ecosystem matures, our proposal for auditing (and more generally, AI auditing as a practice) is only one piece in the responsible AI puzzle, and---much similar to financial auditing---is not the answer to all problems of accountability.

\section{Ethics Statement}

    \subsubsection{Positionality statement.} All authors of this paper work for a for-profit business that has been using this audit framework to perform bias audits under NYC Local Law 144 of 2021 since 2022. This work is thus drawn on our perspectives (e.g., values, norms, practices, and biases, explicitly and implicitly) (1) as consultants providing advisory services, (2) as auditors conducting algorithm audits for companies using and developing AI, and (3) as researchers of responsible AI and AI governance. More specifically, our discussion and prioritization on benefits, limitations, and implementation challenges draw heavily on our experiences having used this audit framework to perform the bias audits. They are, however, not intended to be comprehensive of all audit stakeholders' perspectives (e.g., the public, audit report readers, regulators, or auditees).

    \subsubsection{Adverse impact statement.} We recognize that there are risks associated with the misuse of our methodology. For example, our criteria for NYC bias audit law was designed assuming certain ways that the AI system is built. While we initially tested the criteria’s applicability using publicly available information about AI hiring systems, we could not have imagined all the possible ways hiring algorithms are designed, developed, deployed, and tested. Uncritically using our criteria to perform audits can thus run into issues concerning adaptability to real-world hiring systems. In addition, our criteria can also be ‘reverse engineered’ by companies to game our audit process, although we expect this risk to be of no immediate cause for concern. More generally, there are also risks of misinterpretation of our analysis. Our operationalizable audit framework can be seen as conveying a false sense of what auditing can do to advance accountability without situating it the appropriate context---i.e., that it is an instrument whose effectiveness requires the existence of a supporting multi-stakeholder ecosystem. We also discuss such issues in section \ref{limitations} and \ref{implementation}.

    \subsubsection{Ethical considerations statement.} Due to the nature of our research, we did not undergo any Institutional Review Board (IRB) process. However, as we acknowledge our position as one where audits being widely adopted is a business interest, our discussion is intended to only critically reflect on the ways auditing may benefit the current discourse on AI transparency and accountability.

\begin{acks}

    We would like to thank the Notre Dame-IBM Tech Ethics Lab for supporting this work. Notre Dame-IBM Tech Ethics Lab had no role in the design and conduct of the study; the access and collection of data; the analysis and interpretation of data; the preparation, review, or approval of the manuscript; or the decision to submit the manuscript for publication.
    
\end{acks}

\bibliographystyle{ACM-Reference-Format}
\bibliography{base}


\begin{thebibliography}{84}


\ifx \showCODEN    \undefined \def \showCODEN     #1{\unskip}     \fi
\ifx \showDOI      \undefined \def \showDOI       #1{#1}\fi
\ifx \showISBNx    \undefined \def \showISBNx     #1{\unskip}     \fi
\ifx \showISBNxiii \undefined \def \showISBNxiii  #1{\unskip}     \fi
\ifx \showISSN     \undefined \def \showISSN      #1{\unskip}     \fi
\ifx \showLCCN     \undefined \def \showLCCN      #1{\unskip}     \fi
\ifx \shownote     \undefined \def \shownote      #1{#1}          \fi
\ifx \showarticletitle \undefined \def \showarticletitle #1{#1}   \fi
\ifx \showURL      \undefined \def \showURL       {\relax}        \fi
\providecommand\bibfield[2]{#2}
\providecommand\bibinfo[2]{#2}
\providecommand\natexlab[1]{#1}
\providecommand\showeprint[2][]{arXiv:#2}

\bibitem[{Ada Lovelace Institute} and {DataKind UK}(2020)]%
        {adalovelace}
\bibfield{author}{\bibinfo{person}{{Ada Lovelace Institute}} {and} \bibinfo{person}{{DataKind UK}}.} \bibinfo{year}{2020}\natexlab{}.
\newblock \bibinfo{booktitle}{\emph{Examining the Black Box: Tools for Assessing Algorithmic Systems}}.
\newblock \bibinfo{type}{{T}echnical {R}eport}.
\newblock
\urldef\tempurl%
\url{https://www.adalovelaceinstitute.org/report/examining-the-black-box-tools-for-assessing-algorithmic-systems/}
\showURL{%
\tempurl}


\bibitem[Ajunwa(2021)]%
        {ajunwa_21}
\bibfield{author}{\bibinfo{person}{Ifeoma Ajunwa}.} \bibinfo{year}{2021}\natexlab{}.
\newblock \showarticletitle{An Auditing Imperative for Automated Hiring}.
\newblock \bibinfo{journal}{\emph{Harvard Journal of Law \& Technology}} \bibinfo{volume}{34}, \bibinfo{number}{2} (\bibinfo{date}{06} \bibinfo{year}{2021}), \bibinfo{numpages}{80}~pages.
\newblock
\urldef\tempurl%
\url{https://doi.org/10.2139/ssrn.3437631}
\showDOI{\tempurl}


\bibitem[Angwin and Mattu(2016)]%
        {angwin_16}
\bibfield{author}{\bibinfo{person}{Julia Angwin} {and} \bibinfo{person}{Surya Mattu}.} \bibinfo{year}{2016}\natexlab{}.
\newblock \showarticletitle{Amazon Says It Puts Customers First. But Its Pricing Algorithm Doesn’t}.
\newblock \bibinfo{journal}{\emph{ProPublica}} (\bibinfo{date}{09} \bibinfo{year}{2016}).
\newblock
\urldef\tempurl%
\url{https://www.propublica.org/article/amazon-says-it-puts-customers-first-but-its-pricing-algorithm-doesnt}
\showURL{%
\tempurl}


\bibitem[Barocas and Selbst(2016)]%
        {barocas_16}
\bibfield{author}{\bibinfo{person}{Solon Barocas} {and} \bibinfo{person}{Andrew~D. Selbst}.} \bibinfo{year}{2016}\natexlab{}.
\newblock \showarticletitle{Big Data's Disparate Impact}.
\newblock \bibinfo{journal}{\emph{California Law Review}} \bibinfo{volume}{104}, \bibinfo{number}{3} (\bibinfo{date}{08} \bibinfo{year}{2016}), \bibinfo{pages}{671--732}.
\newblock
\urldef\tempurl%
\url{https://doi.org/10.2139/ssrn.2477899}
\showDOI{\tempurl}


\bibitem[Baxter(2021)]%
        {baxter_21}
\bibfield{author}{\bibinfo{person}{Kathy Baxter}.} \bibinfo{year}{2021}\natexlab{}.
\newblock \bibinfo{booktitle}{\emph{AI Ethics Maturity Model}}.
\newblock \bibinfo{type}{{T}echnical {R}eport}. \bibinfo{institution}{Salesforce AI Research}.
\newblock
\urldef\tempurl%
\url{https://www.salesforceairesearch.com/static/ethics/EthicalAIMaturityModel.pdf}
\showURL{%
\tempurl}


\bibitem[Bell et~al\mbox{.}(2023)]%
        {bell_23}
\bibfield{author}{\bibinfo{person}{Andrew Bell}, \bibinfo{person}{Lucius Bynum}, \bibinfo{person}{Nazarii Drushchak}, \bibinfo{person}{Tetiana Zakharchenko}, \bibinfo{person}{Lucas Rosenblatt}, {and} \bibinfo{person}{Julia Stoyanovich}.} \bibinfo{year}{2023}\natexlab{}.
\newblock \showarticletitle{The Possibility of Fairness: Revisiting the Impossibility Theorem in Practice}. In \bibinfo{booktitle}{\emph{Proceedings of the 2023 ACM Conference on Fairness, Accountability, and Transparency}} (Chicago, IL, USA) \emph{(\bibinfo{series}{FAccT '23})}. \bibinfo{publisher}{Association for Computing Machinery}, \bibinfo{address}{New York, NY, USA}, \bibinfo{pages}{400--422}.
\newblock
\urldef\tempurl%
\url{https://doi.org/10.1145/3593013.3594007}
\showDOI{\tempurl}


\bibitem[Biagioli(2016)]%
        {biagioli_16}
\bibfield{author}{\bibinfo{person}{Mario Biagioli}.} \bibinfo{year}{2016}\natexlab{}.
\newblock \showarticletitle{Watch out for Cheats in Citation Game}.
\newblock \bibinfo{journal}{\emph{Nature}}  \bibinfo{volume}{535} (\bibinfo{date}{07} \bibinfo{year}{2016}), \bibinfo{pages}{201}.
\newblock
\urldef\tempurl%
\url{https://doi.org/10.1038/535201a}
\showDOI{\tempurl}


\bibitem[Binns(2020)]%
        {binns_20}
\bibfield{author}{\bibinfo{person}{Reuben Binns}.} \bibinfo{year}{2020}\natexlab{}.
\newblock \showarticletitle{On the Apparent Conflict between Individual and Group Fairness}. In \bibinfo{booktitle}{\emph{Proceedings of the 2020 Conference on Fairness, Accountability, and Transparency}} (Barcelona, Spain) \emph{(\bibinfo{series}{FAT* '20})}. \bibinfo{publisher}{Association for Computing Machinery}, \bibinfo{address}{New York, NY, USA}, \bibinfo{pages}{514--524}.
\newblock
\urldef\tempurl%
\url{https://doi.org/10.1145/3351095.3372864}
\showDOI{\tempurl}


\bibitem[{Board of Governors of the Federal Reserve System}(2011)]%
        {sr117}
\bibfield{author}{\bibinfo{person}{{Board of Governors of the Federal Reserve System}}.} \bibinfo{year}{2011}\natexlab{}.
\newblock \bibinfo{title}{SR 11-7: Guidance on Model Risk Management}.
\newblock
\newblock
\urldef\tempurl%
\url{https://www.federalreserve.gov/supervisionreg/srletters/sr1107.htm}
\showURL{%
\tempurl}


\bibitem[Brown et~al\mbox{.}(2021)]%
        {brown_21}
\bibfield{author}{\bibinfo{person}{Shea Brown}, \bibinfo{person}{Jovana Davidovic}, {and} \bibinfo{person}{Ali Hasan}.} \bibinfo{year}{2021}\natexlab{}.
\newblock \showarticletitle{The Algorithm Audit: Scoring the Algorithms That Score Us}.
\newblock \bibinfo{journal}{\emph{Big Data \& Society}} \bibinfo{volume}{8}, \bibinfo{number}{1} (\bibinfo{date}{01} \bibinfo{year}{2021}).
\newblock
\urldef\tempurl%
\url{https://doi.org/10.1177/2053951720983865}
\showDOI{\tempurl}


\bibitem[Brundage et~al\mbox{.}(2020)]%
        {brundage_2020}
\bibfield{author}{\bibinfo{person}{Miles Brundage}, \bibinfo{person}{Shahar Avin}, \bibinfo{person}{Jasmine Wang}, \bibinfo{person}{Haydn Belfield}, \bibinfo{person}{Gretchen Krueger}, \bibinfo{person}{Gillian Hadfield}, \bibinfo{person}{Heidy Khlaaf}, \bibinfo{person}{Jingying Yang}, \bibinfo{person}{Helen Toner}, \bibinfo{person}{Ruth Fong}, \bibinfo{person}{Tegan Maharaj}, \bibinfo{person}{Pang~Wei Koh}, \bibinfo{person}{Sara Hooker}, \bibinfo{person}{Jade Leung}, \bibinfo{person}{Andrew Trask}, \bibinfo{person}{Emma Bluemke}, \bibinfo{person}{Jonathan Lebensold}, \bibinfo{person}{Cullen O'Keefe}, \bibinfo{person}{Mark Koren}, \bibinfo{person}{Th{\'e}o Ryffel}, \bibinfo{person}{JB Rubinovitz}, \bibinfo{person}{Tamay Besiroglu}, \bibinfo{person}{Federica Carugati}, \bibinfo{person}{Jack Clark}, \bibinfo{person}{Peter Eckersley}, \bibinfo{person}{Sarah de Haas}, \bibinfo{person}{Maritza Johnson}, \bibinfo{person}{Ben Laurie}, \bibinfo{person}{Alex Ingerman}, \bibinfo{person}{Igor Krawczuk},
  \bibinfo{person}{Amanda Askell}, \bibinfo{person}{Rosario Cammarota}, \bibinfo{person}{Andrew Lohn}, \bibinfo{person}{David Krueger}, \bibinfo{person}{Charlotte Stix}, \bibinfo{person}{Peter Henderson}, \bibinfo{person}{Logan Graham}, \bibinfo{person}{Carina Prunkl}, \bibinfo{person}{Bianca Martin}, \bibinfo{person}{Elizabeth Seger}, \bibinfo{person}{Noa Zilberman}, \bibinfo{person}{Se{\'a}n~{\'O} h{\'E}igeartaigh}, \bibinfo{person}{Frens Kroeger}, \bibinfo{person}{Girish Sastry}, \bibinfo{person}{Rebecca Kagan}, \bibinfo{person}{Adrian Weller}, \bibinfo{person}{Brian Tse}, \bibinfo{person}{Elizabeth Barnes}, \bibinfo{person}{Allan Dafoe}, \bibinfo{person}{Paul Scharre}, \bibinfo{person}{Ariel Herbert-Voss}, \bibinfo{person}{Martijn Rasser}, \bibinfo{person}{Shagun Sodhani}, \bibinfo{person}{Carrick Flynn}, \bibinfo{person}{Thomas Krendl~Gilbert}, \bibinfo{person}{Lisa Dyer}, \bibinfo{person}{Saif Khan}, \bibinfo{person}{Yoshua Bengio}, {and} \bibinfo{person}{Markus Anderljung}.}
  \bibinfo{year}{2020}\natexlab{}.
\newblock \showarticletitle{Toward Trustworthy AI Development: Mechanisms for Supporting Verifiable Claims}.
\newblock  (\bibinfo{date}{04} \bibinfo{year}{2020}).
\newblock
\showeprint{2004.07213}~[cs.CY]


\bibitem[{Bundesamt f{\"u}r Sicherheit in der Informationstechnik (BSI)}(2022)]%
        {aic4}
\bibfield{author}{\bibinfo{person}{{Bundesamt f{\"u}r Sicherheit in der Informationstechnik (BSI)}}.} \bibinfo{year}{2022}\natexlab{}.
\newblock \bibinfo{title}{AI Cloud Service Compliance Criteria Catalogue (AIC4)}.
\newblock
\newblock
\urldef\tempurl%
\url{https://www.bsi.bund.de/EN/Themen/Unternehmen-und-Organisationen/Informationen-und-Empfehlungen/Kuenstliche-Intelligenz/AIC4/aic4_node.html}
\showURL{%
\tempurl}


\bibitem[Buolamwini and Gebru(2018)]%
        {buolamwini_18}
\bibfield{author}{\bibinfo{person}{Joy Buolamwini} {and} \bibinfo{person}{Timnit Gebru}.} \bibinfo{year}{2018}\natexlab{}.
\newblock \showarticletitle{Gender Shades: Intersectional Accuracy Disparities in Commercial Gender Classification}. In \bibinfo{booktitle}{\emph{Proceedings of the 1st Conference on Fairness, Accountability and Transparency}} \emph{(\bibinfo{series}{PMLR}, Vol.~\bibinfo{volume}{81})}, \bibfield{editor}{\bibinfo{person}{Sorelle~A. Friedler} {and} \bibinfo{person}{Christo Wilson}} (Eds.). \bibinfo{publisher}{Proceedings of Machine Learning Research}, \bibinfo{address}{New York, NY, USA}, \bibinfo{pages}{77--91}.
\newblock
\urldef\tempurl%
\url{https://proceedings.mlr.press/v81/buolamwini18a.html}
\showURL{%
\tempurl}


\bibitem[Carrier et~al\mbox{.}(2021)]%
        {fh_nyc}
\bibfield{author}{\bibinfo{person}{Ryan Carrier}, \bibinfo{person}{Shea Brown}, \bibinfo{person}{Merve Hickok}, \bibinfo{person}{Cari Miller}, \bibinfo{person}{Michael McCarthy}, \bibinfo{person}{Esther Chung}, \bibinfo{person}{Joshua Scarpino}, \bibinfo{person}{Heidi Saas}, {and} \bibinfo{person}{Marc H{\'e}bert}.} \bibinfo{year}{2021}\natexlab{}.
\newblock \bibinfo{booktitle}{\emph{New York City Local Law 144: Bias Audits for Automated Employment Decision Tools}}.
\newblock \bibinfo{type}{{T}echnical {R}eport}. \bibinfo{institution}{ForHumanity}.
\newblock
\urldef\tempurl%
\url{https://forhumanity.center/web/wp-content/uploads/2023/10/New-York-City-Bias-Audit-An-Overview-and-Action-Plan-v2.pdf}
\showURL{%
\tempurl}


\bibitem[Choi et~al\mbox{.}(2018)]%
        {choi_18}
\bibfield{author}{\bibinfo{person}{Jong-Hag Choi}, \bibinfo{person}{Heesun Chung}, \bibinfo{person}{Catherine~Heyjung Sonu}, {and} \bibinfo{person}{Yoonseok Zang}.} \bibinfo{year}{2018}\natexlab{}.
\newblock \showarticletitle{Opinion Shopping to Avoid a Going Concern Audit Opinion and Subsequent Audit Quality}.
\newblock \bibinfo{journal}{\emph{Auditing: A Journal of Practice \& Theory}} \bibinfo{volume}{38}, \bibinfo{number}{2} (\bibinfo{date}{05} \bibinfo{year}{2018}), \bibinfo{pages}{101--123}.
\newblock
\urldef\tempurl%
\url{https://papers.ssrn.com/sol3/papers.cfm?abstract_id=3182103}
\showURL{%
\tempurl}


\bibitem[Chouldechova et~al\mbox{.}(2018)]%
        {chouldechova_18}
\bibfield{author}{\bibinfo{person}{Alexandra Chouldechova}, \bibinfo{person}{Diana Benavides-Prado}, \bibinfo{person}{Oleksandr Fialko}, {and} \bibinfo{person}{Rhema Vaithianathan}.} \bibinfo{year}{2018}\natexlab{}.
\newblock \showarticletitle{A Case Study of Algorithm-Assisted Decision Making in Child Maltreatment Hotline Screening Decisions}. In \bibinfo{booktitle}{\emph{Proceedings of the 1st Conference on Fairness, Accountability and Transparency}} \emph{(\bibinfo{series}{PMLR}, Vol.~\bibinfo{volume}{81})}, \bibfield{editor}{\bibinfo{person}{Sorelle~A. Friedler} {and} \bibinfo{person}{Christo Wilson}} (Eds.). \bibinfo{publisher}{Proceedings of Machine Learning Research}, \bibinfo{address}{New York, NY, USA}, \bibinfo{pages}{134--148}.
\newblock
\urldef\tempurl%
\url{https://proceedings.mlr.press/v81/chouldechova18a.html}
\showURL{%
\tempurl}


\bibitem[Clarke(2022)]%
        {algoaccact}
\bibfield{author}{\bibinfo{person}{Yvette~D. Clarke}.} \bibinfo{year}{2022}\natexlab{}.
\newblock \bibinfo{title}{H.R. 6580 -- Algorithmic Accountability Act of 2022}.
\newblock
\newblock
\urldef\tempurl%
\url{https://www.congress.gov/bill/117th-congress/house-bill/6580}
\showURL{%
\tempurl}


\bibitem[Costanza-Chock et~al\mbox{.}(2022)]%
        {costanzachock_22}
\bibfield{author}{\bibinfo{person}{Sasha Costanza-Chock}, \bibinfo{person}{Inioluwa~Deborah Raji}, {and} \bibinfo{person}{Joy Buolamwini}.} \bibinfo{year}{2022}\natexlab{}.
\newblock \showarticletitle{Who Audits the Auditors? Recommendations from a Field Scan of the Algorithmic Auditing Ecosystem}. In \bibinfo{booktitle}{\emph{Proceedings of the 2022 ACM Conference on Fairness, Accountability, and Transparency}} (Seoul, Republic of Korea) \emph{(\bibinfo{series}{FAccT '22})}. \bibinfo{publisher}{Association for Computing Machinery}, \bibinfo{address}{New York, NY, USA}, \bibinfo{pages}{1571--1583}.
\newblock
\urldef\tempurl%
\url{https://doi.org/10.1145/3531146.3533213}
\showDOI{\tempurl}


\bibitem[Dan\'{i}elsson(2002)]%
        {danielsson_02}
\bibfield{author}{\bibinfo{person}{J\'{o}n Dan\'{i}elsson}.} \bibinfo{year}{2002}\natexlab{}.
\newblock \showarticletitle{The Emperor Has No clothes: Limits to Risk Modelling}.
\newblock \bibinfo{journal}{\emph{Journal of Banking \& Finance}} \bibinfo{volume}{26}, \bibinfo{number}{7} (\bibinfo{date}{07} \bibinfo{year}{2002}), \bibinfo{pages}{1273--1296}.
\newblock
\urldef\tempurl%
\url{https://doi.org/10.1016/s0378-4266(02)00263-7}
\showDOI{\tempurl}


\bibitem[{Eightfold AI}(2023)]%
        {8f_23}
\bibfield{author}{\bibinfo{person}{{Eightfold AI}}.} \bibinfo{year}{2023}\natexlab{}.
\newblock \bibinfo{title}{Summary of Bias Audit Results: Audit of Eightfold’s Matching Model for New York City's Local Law 144}.
\newblock
\newblock
\urldef\tempurl%
\url{https://perma.cc/3JGK-7H76}
\showURL{%
\tempurl}


\bibitem[Engler(2021)]%
        {engler_21}
\bibfield{author}{\bibinfo{person}{Alex Engler}.} \bibinfo{year}{2021}\natexlab{}.
\newblock \showarticletitle{Auditing Employment Algorithms for Discrimination}.
\newblock \bibinfo{journal}{\emph{The Brookings Institution}} (\bibinfo{date}{03} \bibinfo{year}{2021}).
\newblock
\urldef\tempurl%
\url{https://www.brookings.edu/articles/auditing-employment-algorithms-for-discrimination/}
\showURL{%
\tempurl}


\bibitem[{European Commission}(2021)]%
        {aiact}
\bibfield{author}{\bibinfo{person}{{European Commission}}.} \bibinfo{year}{2021}\natexlab{}.
\newblock \bibinfo{title}{Proposal for a Regulation of the European Parliament and the Council Laying Down Harmonised Rules on Artificial Intelligence (Artificial Intelligence Act) and Amending Certain Union Legislative Acts}.
\newblock
\newblock
\urldef\tempurl%
\url{https://assets-global.website-files.com/637e4725db842e4068de0899/6565f4809623754436366a2b_COMMISSION%20PROPOSAL.pdf}
\showURL{%
\tempurl}


\bibitem[{European Parliament} and {Council of the European Union}(2022)]%
        {dsa}
\bibfield{author}{\bibinfo{person}{{European Parliament}} {and} \bibinfo{person}{{Council of the European Union}}.} \bibinfo{year}{2022}\natexlab{}.
\newblock \bibinfo{title}{Regulation (EU) 2022/2065 of the European Parliament and of the Council of 19 October 2022 on a Single Market for Digital Services and Amending Directive 2000/31/EC (Digital Services Act)}.
\newblock
\newblock
\urldef\tempurl%
\url{https://eur-lex.europa.eu/legal-content/EN/TXT/?uri=celex%3A32022R2065}
\showURL{%
\tempurl}


\bibitem[{European Parliament} and {Council of the European Union}(2023)]%
        {dsa_audit}
\bibfield{author}{\bibinfo{person}{{European Parliament}} {and} \bibinfo{person}{{Council of the European Union}}.} \bibinfo{year}{2023}\natexlab{}.
\newblock \bibinfo{title}{Delegated Regulation on Independent Audits under the Digital Services Act}.
\newblock
\newblock
\urldef\tempurl%
\url{https://digital-strategy.ec.europa.eu/en/library/delegated-regulation-independent-audits-under-digital-services-act}
\showURL{%
\tempurl}


\bibitem[{Federal Trade Commission}(2022)]%
        {ftc}
\bibfield{author}{\bibinfo{person}{{Federal Trade Commission}}.} \bibinfo{year}{2022}\natexlab{}.
\newblock \bibinfo{title}{Trade Regulation Rule on Commercial Surveillance and Data Security}.
\newblock
\newblock
\urldef\tempurl%
\url{https://www.federalregister.gov/documents/2022/08/22/2022-17752/trade-regulation-rule-on-commercial-surveillance-and-data-security}
\showURL{%
\tempurl}


\bibitem[{ForHumanity}(2022)]%
        {fh_uni}
\bibfield{author}{\bibinfo{person}{{ForHumanity}}.} \bibinfo{year}{2022}\natexlab{}.
\newblock \bibinfo{title}{ForHumanity University}.
\newblock
\newblock
\urldef\tempurl%
\url{https://forhumanity.center/forhumanity-university/}
\showURL{%
\tempurl}


\bibitem[{ForHumanity}(2023)]%
        {fh_manual}
\bibfield{author}{\bibinfo{person}{{ForHumanity}}.} \bibinfo{year}{2023}\natexlab{}.
\newblock \bibinfo{booktitle}{\emph{Audit Manual for Independent Audit of AI Systems v1.5}}.
\newblock
\urldef\tempurl%
\url{https://forhumanity.center/web/wp-content/uploads/2023/08/ForHumanity-IAAIS-Audit-Manual-v1.5.pdf}
\showURL{%
\tempurl}


\bibitem[Goodman and Trehu(2023)]%
        {goodman_23}
\bibfield{author}{\bibinfo{person}{Ellen~P. Goodman} {and} \bibinfo{person}{Julia Trehu}.} \bibinfo{year}{2023}\natexlab{}.
\newblock \showarticletitle{Algorithmic Auditing: Chasing AI Accountability}.
\newblock \bibinfo{journal}{\emph{Santa Clara High Technology Law Journal}} \bibinfo{volume}{39}, \bibinfo{number}{3} (\bibinfo{date}{05} \bibinfo{year}{2023}), \bibinfo{pages}{289--338}.
\newblock
\urldef\tempurl%
\url{https://digitalcommons.law.scu.edu/chtlj/vol39/iss3/1}
\showURL{%
\tempurl}


\bibitem[Groves et~al\mbox{.}(2024)]%
        {groves_24}
\bibfield{author}{\bibinfo{person}{Lara Groves}, \bibinfo{person}{Jacob Metcalf}, \bibinfo{person}{Alayna Kennedy}, \bibinfo{person}{Briana Vecchione}, {and} \bibinfo{person}{Andrew Strait}.} \bibinfo{year}{2024}\natexlab{}.
\newblock \bibinfo{title}{Auditing Work: Exploring the New York City Algorithmic Bias Audit Regime}.
\newblock
\newblock
\showeprint{2402.08101}~[cs.CY]


\bibitem[Guszcza et~al\mbox{.}(2018)]%
        {guszcza_18}
\bibfield{author}{\bibinfo{person}{James Guszcza}, \bibinfo{person}{Iyad Rahwan}, \bibinfo{person}{Will Bible}, \bibinfo{person}{Manuel Cebrian}, {and} \bibinfo{person}{Vic Katyal}.} \bibinfo{year}{2018}\natexlab{}.
\newblock \showarticletitle{Why We Need to Audit Algorithms}.
\newblock \bibinfo{journal}{\emph{Harvard Business Review}} (\bibinfo{date}{11} \bibinfo{year}{2018}).
\newblock
\urldef\tempurl%
\url{https://hbr.org/2018/11/why-we-need-to-audit-algorithms}
\showURL{%
\tempurl}


\bibitem[Hasan et~al\mbox{.}(2022)]%
        {hasan_22}
\bibfield{author}{\bibinfo{person}{Ali Hasan}, \bibinfo{person}{Shea Brown}, \bibinfo{person}{Jovana Davidovic}, \bibinfo{person}{Benjamin Lange}, {and} \bibinfo{person}{Mitt Regan}.} \bibinfo{year}{2022}\natexlab{}.
\newblock \showarticletitle{Algorithmic Bias and Risk Assessments: Lessons from Practice}.
\newblock \bibinfo{journal}{\emph{Digital Society}} \bibinfo{volume}{1}, \bibinfo{number}{2} (\bibinfo{date}{08} \bibinfo{year}{2022}), \bibinfo{numpages}{20}~pages.
\newblock
\urldef\tempurl%
\url{https://doi.org/10.1007/s44206-022-00017-z}
\showDOI{\tempurl}


\bibitem[Hauer et~al\mbox{.}(2023)]%
        {hauer_23}
\bibfield{author}{\bibinfo{person}{Marc~P. Hauer}, \bibinfo{person}{Lena Müller-Kress}, \bibinfo{person}{Gertraud Leimüller}, {and} \bibinfo{person}{Katharina Zweig}.} \bibinfo{year}{2023}\natexlab{}.
\newblock \showarticletitle{Using Assurance Cases to Assure the Fulfillment of Non-functional Requirements of AI-based Systems -- Lessons Learned}. In \bibinfo{booktitle}{\emph{2023 IEEE International Conference on Software Testing, Verification and Validation Workshops (ICSTW)}}. \bibinfo{publisher}{Institute of Electrical and Electronics Engineers (IEEE)}, \bibinfo{address}{Dublin, Ireland}, \bibinfo{pages}{172--179}.
\newblock
\urldef\tempurl%
\url{https://doi.org/10.1109/ICSTW58534.2023.00040}
\showDOI{\tempurl}


\bibitem[{Institute of Electrical and Electronics Engineers (IEEE)}(2021)]%
        {ieee}
\bibfield{author}{\bibinfo{person}{{Institute of Electrical and Electronics Engineers (IEEE)}}.} \bibinfo{year}{2021}\natexlab{}.
\newblock \showarticletitle{IEEE Ontological Standard for Ethically Driven Robotics and Automation Systems}.
\newblock \bibinfo{journal}{\emph{IEEE Std 7007-2021}} (\bibinfo{year}{2021}), \bibinfo{pages}{1--119}.
\newblock
\urldef\tempurl%
\url{https://doi.org/10.1109/IEEESTD.2021.9611206}
\showDOI{\tempurl}


\bibitem[{Institute of Electrical and Electronics Engineers (IEEE)}(2023)]%
        {certaied}
\bibfield{author}{\bibinfo{person}{{Institute of Electrical and Electronics Engineers (IEEE)}}.} \bibinfo{year}{2023}\natexlab{}.
\newblock \bibinfo{title}{IEEE CertifAIEd Authorized Assessor Training}.
\newblock
\newblock
\urldef\tempurl%
\url{https://www.ethosai.ai/home-old-bj5}
\showURL{%
\tempurl}


\bibitem[{International Organization for Standardization (ISO)} and {International Electrotechnical Commission (IEC)}(2023)]%
        {iso42001}
\bibfield{author}{\bibinfo{person}{{International Organization for Standardization (ISO)}} {and} \bibinfo{person}{{International Electrotechnical Commission (IEC)}}.} \bibinfo{year}{2023}\natexlab{}.
\newblock \bibinfo{booktitle}{\emph{ISO/IEC 42001:2023 -- Information Technology -- Artificial Intelligence -- Management System} (\bibinfo{edition}{1st} ed.)}.
\newblock \bibinfo{type}{International Standard}.
\newblock
\urldef\tempurl%
\url{https://www.iso.org/standard/81230.html}
\showURL{%
\tempurl}


\bibitem[Kim(2017)]%
        {kim_17}
\bibfield{author}{\bibinfo{person}{Pauline Kim}.} \bibinfo{year}{2017}\natexlab{}.
\newblock \showarticletitle{Auditing Algorithms for Discrimination}.
\newblock \bibinfo{journal}{\emph{University of Pennsylvania Law Review Online}} \bibinfo{volume}{166}, \bibinfo{number}{17-12-03} (\bibinfo{date}{12} \bibinfo{year}{2017}), \bibinfo{pages}{189--203}.
\newblock
\urldef\tempurl%
\url{https://papers.ssrn.com/sol3/papers.cfm?abstract_id=3093982}
\showURL{%
\tempurl}


\bibitem[Kleinberg et~al\mbox{.}(2017)]%
        {kleinberg_17}
\bibfield{author}{\bibinfo{person}{Jon Kleinberg}, \bibinfo{person}{Sendhil Mullainathan}, {and} \bibinfo{person}{Manish Raghavan}.} \bibinfo{year}{2017}\natexlab{}.
\newblock \showarticletitle{Inherent Trade-Offs in the Fair Determination of Risk Scores}. In \bibinfo{booktitle}{\emph{8th Innovations in Theoretical Computer Science Conference (ITCS 2017)}} \emph{(\bibinfo{series}{Leibniz International Proceedings in Informatics (LIPIcs)}, Vol.~\bibinfo{volume}{67})}, \bibfield{editor}{\bibinfo{person}{Christos~H. Papadimitriou}} (Ed.). \bibinfo{publisher}{Schloss Dagstuhl -- Leibniz-Zentrum f{\"u}r Informatik}, \bibinfo{address}{Dagstuhl, Germany}, \bibinfo{pages}{43:1--43:23}.
\newblock
\urldef\tempurl%
\url{https://doi.org/10.4230/LIPIcs.ITCS.2017.43}
\showDOI{\tempurl}


\bibitem[Larson et~al\mbox{.}(2016)]%
        {larson_16}
\bibfield{author}{\bibinfo{person}{Jeff Larson}, \bibinfo{person}{Surya Mattu}, \bibinfo{person}{Lauren Kirchner}, {and} \bibinfo{person}{Julia Angwin}.} \bibinfo{year}{2016}\natexlab{}.
\newblock \showarticletitle{How We Analyzed the COMPAS Recidivism Algorithm}.
\newblock \bibinfo{journal}{\emph{ProPublica}} (\bibinfo{date}{05} \bibinfo{year}{2016}).
\newblock
\urldef\tempurl%
\url{https://www.propublica.org/article/how-we-analyzed-the-compas-recidivism-algorithm}
\showURL{%
\tempurl}


\bibitem[Lennox(2000)]%
        {lennox_00}
\bibfield{author}{\bibinfo{person}{Clive Lennox}.} \bibinfo{year}{2000}\natexlab{}.
\newblock \showarticletitle{Do Companies Successfully Engage in Opinion-shopping? Evidence from the UK}.
\newblock \bibinfo{journal}{\emph{Journal of Accounting and Economics}} \bibinfo{volume}{29}, \bibinfo{number}{3} (\bibinfo{date}{06} \bibinfo{year}{2000}), \bibinfo{pages}{321--337}.
\newblock
\urldef\tempurl%
\url{https://doi.org/10.1016/s0165-4101(00)00025-2}
\showDOI{\tempurl}


\bibitem[Lucaj et~al\mbox{.}(2023)]%
        {lucaj_23}
\bibfield{author}{\bibinfo{person}{Laura Lucaj}, \bibinfo{person}{Patrick van~der Smagt}, {and} \bibinfo{person}{Djalel Benbouzid}.} \bibinfo{year}{2023}\natexlab{}.
\newblock \showarticletitle{AI Regulation Is (Not) All You Need}. In \bibinfo{booktitle}{\emph{Proceedings of the 2023 ACM Conference on Fairness, Accountability, and Transparency}} (Chicago, IL, USA) \emph{(\bibinfo{series}{FAccT '23})}. \bibinfo{publisher}{Association for Computing Machinery}, \bibinfo{address}{New York, NY, USA}, \bibinfo{pages}{1267--1279}.
\newblock
\urldef\tempurl%
\url{https://doi.org/10.1145/3593013.3594079}
\showDOI{\tempurl}


\bibitem[McCann(2019)]%
        {mccann_19}
\bibfield{author}{\bibinfo{person}{David McCann}.} \bibinfo{year}{2019}\natexlab{}.
\newblock \showarticletitle{‘Opinion-Shopping’ Compromises Auditor Independence}.
\newblock \bibinfo{journal}{\emph{CFO.com}} (\bibinfo{date}{05} \bibinfo{year}{2019}).
\newblock
\urldef\tempurl%
\url{https://www.cfo.com/news/opinion-shopping-compromises-auditor-independence/657865/}
\showURL{%
\tempurl}


\bibitem[Mendelson(2021)]%
        {stopdiscact}
\bibfield{author}{\bibinfo{person}{Phil Mendelson}.} \bibinfo{year}{2021}\natexlab{}.
\newblock \bibinfo{title}{B24-0558 -- Stop Discrimination by Algorithms Act of 2021}.
\newblock
\newblock
\urldef\tempurl%
\url{https://legiscan.com/DC/bill/B24-0558/2021}
\showURL{%
\tempurl}


\bibitem[Metcalf et~al\mbox{.}(2021)]%
        {metcalf_21}
\bibfield{author}{\bibinfo{person}{Jacob Metcalf}, \bibinfo{person}{Emanuel Moss}, \bibinfo{person}{Elizabeth~Anne Watkins}, \bibinfo{person}{Ranjit Singh}, {and} \bibinfo{person}{Madeleine~Clare Elish}.} \bibinfo{year}{2021}\natexlab{}.
\newblock \showarticletitle{Algorithmic Impact Assessments and Accountability: The Co-Construction of Impacts}. In \bibinfo{booktitle}{\emph{Proceedings of the 2021 ACM Conference on Fairness, Accountability, and Transparency}} (Virtual Event, Canada) \emph{(\bibinfo{series}{FAccT '21})}. \bibinfo{publisher}{Association for Computing Machinery}, \bibinfo{address}{New York, NY, USA}, \bibinfo{pages}{735--746}.
\newblock
\urldef\tempurl%
\url{https://doi.org/10.1145/3442188.3445935}
\showDOI{\tempurl}


\bibitem[M{\"o}kander(2023)]%
        {mokander_23_ds}
\bibfield{author}{\bibinfo{person}{Jakob M{\"o}kander}.} \bibinfo{year}{2023}\natexlab{}.
\newblock \showarticletitle{Auditing of AI: Legal, Ethical and Technical Approaches}.
\newblock \bibinfo{journal}{\emph{Digital Society}} \bibinfo{volume}{2}, \bibinfo{number}{49} (\bibinfo{year}{2023}), \bibinfo{numpages}{32}~pages.
\newblock
\urldef\tempurl%
\url{https://doi.org/10.1007/s44206-023-00074-y}
\showDOI{\tempurl}


\bibitem[M{\"o}kander et~al\mbox{.}(2022)]%
        {mokander_22}
\bibfield{author}{\bibinfo{person}{Jakob M{\"o}kander}, \bibinfo{person}{Maria Axente}, \bibinfo{person}{Federico Casolari}, {and} \bibinfo{person}{Luciano Floridi}.} \bibinfo{year}{2022}\natexlab{}.
\newblock \showarticletitle{Conformity Assessments and Post-market Monitoring: A Guide to the Role of Auditing in the Proposed European AI Regulation}.
\newblock \bibinfo{journal}{\emph{Minds and Machines}}  \bibinfo{volume}{32} (\bibinfo{year}{2022}), \bibinfo{pages}{241--268}.
\newblock
\urldef\tempurl%
\url{https://doi.org/10.1007/s11023-021-09577-4}
\showDOI{\tempurl}


\bibitem[M{\"o}kander et~al\mbox{.}(2023)]%
        {mokander_23}
\bibfield{author}{\bibinfo{person}{Jakob M{\"o}kander}, \bibinfo{person}{Jonas Schuett}, \bibinfo{person}{Hannah~Rose Kirk}, {and} \bibinfo{person}{Luciano Floridi}.} \bibinfo{year}{2023}\natexlab{}.
\newblock \showarticletitle{Auditing Large Language Models: A Three-Layered Approach}.
\newblock \bibinfo{journal}{\emph{SSRN Electronic Journal}} (\bibinfo{date}{02} \bibinfo{year}{2023}), \bibinfo{numpages}{29}~pages.
\newblock
\urldef\tempurl%
\url{https://doi.org/10.2139/ssrn.4361607}
\showDOI{\tempurl}


\bibitem[{National Institute of Standards and Technology (NIST)}(2023)]%
        {nist_airmf}
\bibfield{author}{\bibinfo{person}{{National Institute of Standards and Technology (NIST)}}.} \bibinfo{year}{2023}\natexlab{}.
\newblock \bibinfo{booktitle}{\emph{Artificial Intelligence Risk Management Framework (AI RMF 1.0)}}.
\newblock \bibinfo{type}{NIST AI} 100-1.
\newblock
\urldef\tempurl%
\url{https://doi.org/10.6028/nist.ai.100-1}
\showDOI{\tempurl}


\bibitem[{NYC Department of Consumer and Worker Protection (DCWP)}(2023)]%
        {nyc_faq}
\bibfield{author}{\bibinfo{person}{{NYC Department of Consumer and Worker Protection (DCWP)}}.} \bibinfo{year}{2023}\natexlab{}.
\newblock \bibinfo{title}{Automated Employment Decision Tools: Frequently Asked Questions}.
\newblock
\newblock
\urldef\tempurl%
\url{https://www.nyc.gov/assets/dca/downloads/pdf/about/DCWP-AEDT-FAQ.pdf}
\showURL{%
\tempurl}


\bibitem[{Office of Federal Contract Compliance Programs (OFCCP)}(1978)]%
        {disp}
\bibfield{author}{\bibinfo{person}{{Office of Federal Contract Compliance Programs (OFCCP)}}.} \bibinfo{year}{1978}\natexlab{}.
\newblock \bibinfo{title}{41 CFR 60-3.4: Information on Impact}.
\newblock
\newblock
\urldef\tempurl%
\url{https://www.ecfr.gov/current/title-41/part-60-3/section-60-3.4}
\showURL{%
\tempurl}


\bibitem[{Office of Federal Contract Compliance Programs (OFCCP)}(2019)]%
        {ugesp}
\bibfield{author}{\bibinfo{person}{{Office of Federal Contract Compliance Programs (OFCCP)}}.} \bibinfo{year}{2019}\natexlab{}.
\newblock \bibinfo{title}{Validation of Employee Selection Procedures: How does OFCCP Identify Disparities (Adverse Impact) Caused by Use of Employee Selection Procedures?}
\newblock
\newblock
\urldef\tempurl%
\url{https://www.dol.gov/agencies/ofccp/faqs/employee-selection-procedures#Q4}
\showURL{%
\tempurl}


\bibitem[{Office of the Comptroller of the Currency (OCC)}(2021)]%
        {occ}
\bibfield{author}{\bibinfo{person}{{Office of the Comptroller of the Currency (OCC)}}.} \bibinfo{year}{2021}\natexlab{}.
\newblock \bibinfo{booktitle}{\emph{Model Risk Management}}.
\newblock \bibinfo{type}{Comptroller's Handbook}.
\newblock
\urldef\tempurl%
\url{https://www.occ.gov/publications-and-resources/publications/comptrollers-handbook/files/model-risk-management/index-model-risk-management.html}
\showURL{%
\tempurl}


\bibitem[Pallone(2022)]%
        {amerprivact}
\bibfield{author}{\bibinfo{person}{Frank Pallone}.} \bibinfo{year}{2022}\natexlab{}.
\newblock \bibinfo{title}{H.R.8152 -- American Data Privacy and Protection Act}.
\newblock
\newblock
\urldef\tempurl%
\url{https://www.congress.gov/bill/117th-congress/house-bill/8152}
\showURL{%
\tempurl}


\bibitem[Power(1994)]%
        {power_94}
\bibfield{author}{\bibinfo{person}{Michael Power}.} \bibinfo{year}{1994}\natexlab{}.
\newblock \bibinfo{booktitle}{\emph{The Audit Explosion}}.
\newblock \bibinfo{type}{{T}echnical {R}eport}. \bibinfo{institution}{Demos}, \bibinfo{address}{London, UK}.
\newblock
\urldef\tempurl%
\url{https://www.demos.co.uk/files/theauditexplosion.pdf}
\showURL{%
\tempurl}


\bibitem[Power(1999)]%
        {power_99}
\bibfield{author}{\bibinfo{person}{Michael Power}.} \bibinfo{year}{1999}\natexlab{}.
\newblock \bibinfo{booktitle}{\emph{The Audit Society: Rituals of Verification}}.
\newblock \bibinfo{publisher}{Oxford University Press}, \bibinfo{address}{Oxford, UK}.
\newblock
\urldef\tempurl%
\url{https://doi.org/10.1093/acprof:oso/9780198296034.001.0001}
\showDOI{\tempurl}


\bibitem[{Public Company Accounting Oversight Board (PCAOB)}(2022)]%
        {as1105}
\bibfield{author}{\bibinfo{person}{{Public Company Accounting Oversight Board (PCAOB)}}.} \bibinfo{year}{2022}\natexlab{}.
\newblock \bibinfo{title}{AS 1105: Audit Evidence}.
\newblock
\newblock
\urldef\tempurl%
\url{https://pcaobus.org/oversight/standards/auditing-standards/details/AS1105}
\showURL{%
\tempurl}


\bibitem[Raji et~al\mbox{.}(2020)]%
        {raji_20}
\bibfield{author}{\bibinfo{person}{Inioluwa~Deborah Raji}, \bibinfo{person}{Andrew Smart}, \bibinfo{person}{Rebecca~N. White}, \bibinfo{person}{Margaret Mitchell}, \bibinfo{person}{Timnit Gebru}, \bibinfo{person}{Ben Hutchinson}, \bibinfo{person}{Jamila Smith-Loud}, \bibinfo{person}{Daniel Theron}, {and} \bibinfo{person}{Parker Barnes}.} \bibinfo{year}{2020}\natexlab{}.
\newblock \showarticletitle{Closing the AI Accountability Gap: Defining an End-to-end Framework for Internal Algorithmic Auditing}. In \bibinfo{booktitle}{\emph{Proceedings of the 2020 Conference on Fairness, Accountability, and Transparency}} (Barcelona, Spain) \emph{(\bibinfo{series}{FAT* '20})}. \bibinfo{publisher}{Association for Computing Machinery}, \bibinfo{address}{New York, NY, USA}, \bibinfo{pages}{33--44}.
\newblock
\urldef\tempurl%
\url{https://doi.org/10.1145/3351095.3372873}
\showDOI{\tempurl}


\bibitem[Raji et~al\mbox{.}(2022)]%
        {raji_22}
\bibfield{author}{\bibinfo{person}{Inioluwa~Deborah Raji}, \bibinfo{person}{Peggy Xu}, \bibinfo{person}{Colleen Honigsberg}, {and} \bibinfo{person}{Daniel Ho}.} \bibinfo{year}{2022}\natexlab{}.
\newblock \showarticletitle{Outsider Oversight: Designing a Third Party Audit Ecosystem for AI Governance}. In \bibinfo{booktitle}{\emph{Proceedings of the 2022 AAAI/ACM Conference on AI, Ethics, and Society}} (Oxford, United Kingdom) \emph{(\bibinfo{series}{AIES '22})}. \bibinfo{publisher}{Association for Computing Machinery}, \bibinfo{address}{New York, NY, USA}, \bibinfo{pages}{557--571}.
\newblock
\urldef\tempurl%
\url{https://doi.org/10.1145/3514094.3534181}
\showDOI{\tempurl}


\bibitem[{RippleMatch}(2023)]%
        {ripple_23}
\bibfield{author}{\bibinfo{person}{{RippleMatch}}.} \bibinfo{year}{2023}\natexlab{}.
\newblock \bibinfo{title}{Summary of Bias Audit Results: Audit of RippleMatch's Fit Score Algorithm for New York City's Local Law 144}.
\newblock
\newblock
\urldef\tempurl%
\url{https://perma.cc/BXW6-7EMA}
\showURL{%
\tempurl}


\bibitem[Sarbanes and Oxley(2002)]%
        {sox}
\bibfield{author}{\bibinfo{person}{Paul Sarbanes} {and} \bibinfo{person}{Michael Oxley}.} \bibinfo{year}{2002}\natexlab{}.
\newblock \bibinfo{title}{Sarbanes-Oxley Act of 2002}.
\newblock
\newblock
\urldef\tempurl%
\url{https://sarbanes-oxley-act.com}
\showURL{%
\tempurl}


\bibitem[Savin-Baden and Howell~Major(2012)]%
        {savinbaden_12}
\bibfield{author}{\bibinfo{person}{Maggi Savin-Baden} {and} \bibinfo{person}{Claire Howell~Major}.} \bibinfo{year}{2012}\natexlab{}.
\newblock \bibinfo{booktitle}{\emph{Qualitative Research: The Essential Guide to Theory and Practice} (\bibinfo{edition}{1st} ed.)}.
\newblock \bibinfo{publisher}{Routledge}, \bibinfo{address}{Oxfordshire, UK}.
\newblock


\bibitem[Schuett(2022)]%
        {schuett_22}
\bibfield{author}{\bibinfo{person}{Jonas Schuett}.} \bibinfo{year}{2022}\natexlab{}.
\newblock \bibinfo{title}{Three Lines of Defense against Risks from AI}.
\newblock , \bibinfo{numpages}{22}~pages.
\newblock
\showeprint{2212.08364}~[cs.CY]


\bibitem[Schwartz et~al\mbox{.}(2022)]%
        {schwartz_22}
\bibfield{author}{\bibinfo{person}{Reva Schwartz}, \bibinfo{person}{Apostol Vassilev}, \bibinfo{person}{Kristen Greene}, \bibinfo{person}{Lori Perine}, \bibinfo{person}{Andrew Burt}, {and} \bibinfo{person}{Patrick Hall}.} \bibinfo{year}{2022}\natexlab{}.
\newblock \bibinfo{booktitle}{\emph{Towards a Standard for Identifying and Managing Bias in Artificial Intelligence}}.
\newblock \bibinfo{type}{NIST Special Publication} 1270. \bibinfo{institution}{National Institute of Standards and Technology (NIST)}.
\newblock
\urldef\tempurl%
\url{https://doi.org/10.6028/nist.sp.1270}
\showDOI{\tempurl}


\bibitem[{Securities and Exchange Commission (SEC)}(1972)]%
        {acc_sec}
\bibfield{author}{\bibinfo{person}{{Securities and Exchange Commission (SEC)}}.} \bibinfo{year}{1972}\natexlab{}.
\newblock \bibinfo{title}{17 CFR 210.2-01: Qualifications of Accountants}.
\newblock
\newblock
\urldef\tempurl%
\url{https://www.ecfr.gov/current/title-17/part-210/section-210.2-01}
\showURL{%
\tempurl}


\bibitem[Selbst(2021)]%
        {selbst_21}
\bibfield{author}{\bibinfo{person}{Andrew~D. Selbst}.} \bibinfo{year}{2021}\natexlab{}.
\newblock \showarticletitle{An Institutional View of Algorithmic Impact Assessments}.
\newblock \bibinfo{journal}{\emph{Harvard Journal of Law \& Technology}} \bibinfo{volume}{35}, \bibinfo{number}{1} (\bibinfo{date}{06} \bibinfo{year}{2021}), \bibinfo{pages}{117--191}.
\newblock
\urldef\tempurl%
\url{https://papers.ssrn.com/abstract=3867634}
\showURL{%
\tempurl}


\bibitem[Selbst et~al\mbox{.}(2019)]%
        {selbst_19}
\bibfield{author}{\bibinfo{person}{Andrew~D. Selbst}, \bibinfo{person}{Danah Boyd}, \bibinfo{person}{Sorelle~A. Friedler}, \bibinfo{person}{Suresh Venkatasubramanian}, {and} \bibinfo{person}{Janet Vertesi}.} \bibinfo{year}{2019}\natexlab{}.
\newblock \showarticletitle{Fairness and Abstraction in Sociotechnical Systems}. In \bibinfo{booktitle}{\emph{Proceedings of the Conference on Fairness, Accountability, and Transparency}} (Atlanta, GA, USA) \emph{(\bibinfo{series}{FAT* '19})}. \bibinfo{publisher}{Association for Computing Machinery}, \bibinfo{address}{New York, NY, USA}, \bibinfo{pages}{59--68}.
\newblock
\urldef\tempurl%
\url{https://doi.org/10.1145/3287560.3287598}
\showDOI{\tempurl}


\bibitem[Sloane(2021)]%
        {sloane_21}
\bibfield{author}{\bibinfo{person}{Mona Sloane}.} \bibinfo{year}{2021}\natexlab{}.
\newblock \showarticletitle{The Algorithmic Auditing Trap}.
\newblock \bibinfo{journal}{\emph{OneZero}} (\bibinfo{date}{03} \bibinfo{year}{2021}).
\newblock
\urldef\tempurl%
\url{https://onezero.medium.com/the-algorithmic-auditing-trap-9a6f2d4d461d}
\showURL{%
\tempurl}


\bibitem[Sloane and Moss(2023)]%
        {sloane_23}
\bibfield{author}{\bibinfo{person}{Mona Sloane} {and} \bibinfo{person}{Emanuel Moss}.} \bibinfo{year}{2023}\natexlab{}.
\newblock \showarticletitle{Assessing the Assessment: Comparing Algorithmic Impact Assessments and AI Audits}.
\newblock \bibinfo{journal}{\emph{SSRN Electronic Journal}} (\bibinfo{date}{06} \bibinfo{year}{2023}), \bibinfo{numpages}{14}~pages.
\newblock
\urldef\tempurl%
\url{https://doi.org/10.2139/ssrn.4486259}
\showDOI{\tempurl}
\newblock
\shownote{In review for edited volume for Oxford University Press}.


\bibitem[Snow(2018)]%
        {snow_18}
\bibfield{author}{\bibinfo{person}{Jacob Snow}.} \bibinfo{year}{2018}\natexlab{}.
\newblock \showarticletitle{Amazon’s Face Recognition Falsely Matched 28 Members of Congress with Mugshots}.
\newblock \bibinfo{journal}{\emph{American Civil Liberties Union}} (\bibinfo{date}{07} \bibinfo{year}{2018}).
\newblock
\urldef\tempurl%
\url{https://www.aclu.org/news/privacy-technology/amazons-face-recognition-falsely-matched-28}
\showURL{%
\tempurl}


\bibitem[{The Algorithmic Bias Lab}(2022)]%
        {bablcert}
\bibfield{author}{\bibinfo{person}{{The Algorithmic Bias Lab}}.} \bibinfo{year}{2022}\natexlab{}.
\newblock \bibinfo{title}{AI and Algorithm Auditor Certificate Program}.
\newblock
\newblock
\urldef\tempurl%
\url{https://courses.babl.ai/p/ai-and-algorithm-auditor-certification}
\showURL{%
\tempurl}


\bibitem[{The International Auditing and Assurance Standards Board (IAASB)}(2013)]%
        {isae3000}
\bibfield{author}{\bibinfo{person}{{The International Auditing and Assurance Standards Board (IAASB)}}.} \bibinfo{year}{2013}\natexlab{}.
\newblock \bibinfo{booktitle}{\emph{International Standard on Assurance Engagements (ISAE) 3000 Revised, Assurance Engagements Other Than Audits or Reviews of Historical Financial Information}}.
\newblock \bibinfo{type}{Standards and Pronouncements} 978-1-60815-167-7. \bibinfo{institution}{The International Federation of Accountants (IFAC)}.
\newblock
\urldef\tempurl%
\url{https://www.iaasb.org/publications/international-standard-assurance-engagements-isae-3000-revised-assurance-engagements-other-audits-or}
\showURL{%
\tempurl}


\bibitem[{The International Auditing and Assurance Standards Board (IAASB)}(2014)]%
        {audit_q}
\bibfield{author}{\bibinfo{person}{{The International Auditing and Assurance Standards Board (IAASB)}}.} \bibinfo{year}{2014}\natexlab{}.
\newblock \bibinfo{booktitle}{\emph{A Framework for Audit Quality: Key Elements That Create an Environment for Audit Quality}}.
\newblock \bibinfo{type}{Standards and Pronouncements} 978-1-60815-178-3. \bibinfo{institution}{The International Federation of Accountants (IFAC)}.
\newblock
\urldef\tempurl%
\url{https://www.iaasb.org/publications/framework-audit-quality-key-elements-create-environment-audit-quality-3}
\showURL{%
\tempurl}


\bibitem[{The International Auditing and Assurance Standards Board (IAASB)}(2019)]%
        {isa315}
\bibfield{author}{\bibinfo{person}{{The International Auditing and Assurance Standards Board (IAASB)}}.} \bibinfo{year}{2019}\natexlab{}.
\newblock \bibinfo{booktitle}{\emph{ISA 315 (Revised 2019): Identifying and Assessing the Risks of Material Misstatement}}.
\newblock \bibinfo{type}{Standards and Pronouncements}. \bibinfo{institution}{The International Federation of Accountants (IFAC)}.
\newblock
\urldef\tempurl%
\url{https://www.iaasb.org/publications/isa-315-revised-2019-identifying-and-assessing-risks-material-misstatement}
\showURL{%
\tempurl}


\bibitem[{The International Auditing and Assurance Standards Board (IAASB)}(2020)]%
        {isqm}
\bibfield{author}{\bibinfo{person}{{The International Auditing and Assurance Standards Board (IAASB)}}.} \bibinfo{year}{2020}\natexlab{}.
\newblock \bibinfo{booktitle}{\emph{International Standard on Quality Management (ISQM) 1: Quality Management for Firms That Perform Audits or Reviews of Financial Statements, or Other Assurance or Related Services Engagements}}.
\newblock \bibinfo{type}{Handbooks, Standards, and Pronouncements}. \bibinfo{institution}{The International Federation of Accountants (IFAC)}.
\newblock
\urldef\tempurl%
\url{https://www.iaasb.org/publications/international-standard-quality-management-isqm-1-quality-management-firms-perform-audits-or-reviews}
\showURL{%
\tempurl}


\bibitem[{The International Auditing and Assurance Standards Board (IAASB)}(2022)]%
        {acc_handbook}
\bibfield{author}{\bibinfo{person}{{The International Auditing and Assurance Standards Board (IAASB)}}.} \bibinfo{year}{2022}\natexlab{}.
\newblock \bibinfo{booktitle}{\emph{2022 Handbook of the International Code Of Ethics for Professional Accountants}}.
\newblock \bibinfo{type}{Handbooks} 978-1-60815-508-8. \bibinfo{institution}{The International Federation of Accountants (IFAC)}.
\newblock
\urldef\tempurl%
\url{https://www.ethicsboard.org/publications/2022-handbook-international-code-ethics-professional-accountants}
\showURL{%
\tempurl}


\bibitem[{The New York City Council}(2021)]%
        {nyc_ll144}
\bibfield{author}{\bibinfo{person}{{The New York City Council}}.} \bibinfo{year}{2021}\natexlab{}.
\newblock \bibinfo{title}{Subchapter 25: Automated Employment Decision Tools}.
\newblock
\newblock
\urldef\tempurl%
\url{https://codelibrary.amlegal.com/codes/newyorkcity/latest/NYCadmin/0-0-0-135598}
\showURL{%
\tempurl}


\bibitem[{The New York City Council}(2023)]%
        {nyc_chapT}
\bibfield{author}{\bibinfo{person}{{The New York City Council}}.} \bibinfo{year}{2023}\natexlab{}.
\newblock \bibinfo{title}{Subchapter T: Automated Employment Decision Tools}.
\newblock
\newblock
\urldef\tempurl%
\url{https://codelibrary.amlegal.com/codes/newyorkcity/latest/NYCrules/0-0-0-138391}
\showURL{%
\tempurl}


\bibitem[{U.S. Department of Health \& Human Services}(2021)]%
        {hhs}
\bibfield{author}{\bibinfo{person}{{U.S. Department of Health \& Human Services}}.} \bibinfo{year}{2021}\natexlab{}.
\newblock \bibinfo{booktitle}{\emph{Trustworthy AI (TAI) Playbook}}.
\newblock \bibinfo{type}{{T}echnical {R}eport}.
\newblock
\urldef\tempurl%
\url{https://www.hhs.gov/sites/default/files/hhs-trustworthy-ai-playbook.pdf}
\showURL{%
\tempurl}


\bibitem[{U.S. Government Accountability Office (GAO)}(2021)]%
        {gao}
\bibfield{author}{\bibinfo{person}{{U.S. Government Accountability Office (GAO)}}.} \bibinfo{year}{2021}\natexlab{}.
\newblock \bibinfo{booktitle}{\emph{Artificial Intelligence: An Accountability Framework for Federal Agencies and Other Entities}}.
\newblock \bibinfo{type}{{T}echnical {R}eport} GAO-21-519SP.
\newblock
\urldef\tempurl%
\url{https://www.gao.gov/products/gao-21-519sp}
\showURL{%
\tempurl}


\bibitem[Wagner and Kuklis(2021)]%
        {wagner_21}
\bibfield{author}{\bibinfo{person}{Ben Wagner} {and} \bibinfo{person}{Lubos Kuklis}.} \bibinfo{year}{2021}\natexlab{}.
\newblock \bibinfo{booktitle}{\emph{Establishing Auditing Intermediaries to Verify Platform Data}}.
\newblock \bibinfo{publisher}{Oxford University Press}, \bibinfo{address}{Oxford, UK}, Chapter~9, \bibinfo{pages}{169--179}.
\newblock
\urldef\tempurl%
\url{https://doi.org/10.1093/oso/9780197616093.003.0010}
\showDOI{\tempurl}


\bibitem[Wall and Schellmann(2021)]%
        {wall_21}
\bibfield{author}{\bibinfo{person}{Sheridan Wall} {and} \bibinfo{person}{Hilke Schellmann}.} \bibinfo{year}{2021}\natexlab{}.
\newblock \showarticletitle{We Tested AI Interview Tools. Here’s What We Found.}
\newblock \bibinfo{journal}{\emph{MIT Technology Review}} (\bibinfo{date}{07} \bibinfo{year}{2021}).
\newblock
\urldef\tempurl%
\url{https://www.technologyreview.com/2021/07/07/1027916/we-tested-ai-interview-tools/}
\showURL{%
\tempurl}


\bibitem[Watkins et~al\mbox{.}(2022)]%
        {watkins_22}
\bibfield{author}{\bibinfo{person}{Elizabeth~Anne Watkins}, \bibinfo{person}{Michael McKenna}, {and} \bibinfo{person}{Jiahao Chen}.} \bibinfo{year}{2022}\natexlab{}.
\newblock \bibinfo{booktitle}{\emph{The Four-Fifths Rule Is Not Disparate Impact: A Woeful Tale of Epistemic Trespassing in Algorithmic Fairness}}.
\newblock \bibinfo{type}{{T}echnical {R}eport} P22-1-v0.2.2. \bibinfo{institution}{Parity Technologies, Inc.}
\newblock
\urldef\tempurl%
\url{https://doi.org/10.48550/arXiv.2202.09519}
\showDOI{\tempurl}


\bibitem[Wilson et~al\mbox{.}(2021)]%
        {wilson_21}
\bibfield{author}{\bibinfo{person}{Christo Wilson}, \bibinfo{person}{Avijit Ghosh}, \bibinfo{person}{Shan Jiang}, \bibinfo{person}{Alan Mislove}, \bibinfo{person}{Lewis Baker}, \bibinfo{person}{Janelle Szary}, \bibinfo{person}{Kelly Trindel}, {and} \bibinfo{person}{Frida Polli}.} \bibinfo{year}{2021}\natexlab{}.
\newblock \showarticletitle{Building and Auditing Fair Algorithms: A Case Study in Candidate Screening}. In \bibinfo{booktitle}{\emph{Proceedings of the 2021 ACM Conference on Fairness, Accountability, and Transparency}} (Virtual Event, Canada) \emph{(\bibinfo{series}{FAccT '21})}. \bibinfo{publisher}{Association for Computing Machinery}, \bibinfo{address}{New York, NY, USA}, \bibinfo{pages}{666--677}.
\newblock
\urldef\tempurl%
\url{https://doi.org/10.1145/3442188.3445928}
\showDOI{\tempurl}


\bibitem[Wright et~al\mbox{.}(2024)]%
        {wright_24}
\bibfield{author}{\bibinfo{person}{Lucas Wright}, \bibinfo{person}{Roxana~Mika Muenster}, \bibinfo{person}{Briana Vecchione}, \bibinfo{person}{Tianyao Qu}, \bibinfo{person}{Pika~(Senhuang) Cai}, \bibinfo{person}{Alan Smith}, \bibinfo{person}{COMM/INFO 2450~STUDENT INVESTIGATORS}, \bibinfo{person}{Jake Metcalf}, {and} \bibinfo{person}{J.~Nathan Matias}.} \bibinfo{year}{2024}\natexlab{}.
\newblock \bibinfo{title}{Null Compliance: NYC Local Law 144 and the Challenges of Algorithm Accountability}.
\newblock
\newblock
\urldef\tempurl%
\url{https://doi.org/10.17605/OSF.IO/UPFDK}
\showDOI{\tempurl}


\bibitem[Young et~al\mbox{.}(2022)]%
        {young_22}
\bibfield{author}{\bibinfo{person}{Meg Young}, \bibinfo{person}{Michael Katell}, {and} \bibinfo{person}{P.~M. Krafft}.} \bibinfo{year}{2022}\natexlab{}.
\newblock \showarticletitle{Confronting Power and Corporate Capture at the FAccT Conference}. In \bibinfo{booktitle}{\emph{2022 ACM Conference on Fairness, Accountability, and Transparency}} (Seoul, Republic of Korea) \emph{(\bibinfo{series}{FAccT '22})}. \bibinfo{publisher}{Association for Computing Machinery}, \bibinfo{address}{New York, NY, USA}, \bibinfo{pages}{1375--1386}.
\newblock
\urldef\tempurl%
\url{https://doi.org/10.1145/3531146.3533194}
\showDOI{\tempurl}


\end{thebibliography}

\appendix

\section{Full Audit Criteria Set} \label{full-criteria}

    Table \ref{tab:full} shows the full set of criteria including all sub-criteria for NYC bias audit law. Note that while the official Frequently Asked Questions (FAQ) by the regulators prohibits inferring demographic labels \cite{nyc_faq}, our interaction with the regulators suggest an exception. During the public Q\&A session and subsequent follow-ups, it came to light that inference may be considered acceptable only as \textit{test data} but not as historical data. We therefore added specific criteria to cover this exception.

    \begin{table*}[ht]
        
        \caption{Full set of audit criteria for NYC Local Law 144 of 2021.}

        \label{tab:full}
        \begin{tabular}{lp{0.8\textwidth}}
            
            \toprule
            ID & \textit{Criterion} \& Sub-criterion \\ 
            
            \midrule
            \textbf{Q} & \multicolumn{1}{c}{\textbf{Disparate Impact Analysis}}\\ 
            
            \midrule
            \textit{Q.A} & \textit{The tool analyzed for disparate impact shall be defined.} \\[\baselineskip]
            Q.A.1 & Where the tool comprises more than one automated component, evidence shall show appropriate definition of the tool. \\
            
            \midrule
            \textit{Q.B} & \textit{The dataset based on which disparate impact is analyzed shall be defined and characterized.} \\[\baselineskip]
            Q.B.1 & Evidence shall show justification for why this dataset is appropriate for analysis. \\
            Q.B.2 & {Where test data as defined in \S 5-300 \cite{nyc_chapT} was used, evidence shall show:
                \begin{enumerate}[label=(\alph*)]
                    \item justification for not using historical data,
                    \item that the sample size of historical data is not sufficiently large to perform a statistically significant disparate impact analysis, and
                    \item the methodology by which the test data was collected.
                \end{enumerate}}\\
            Q.B.3 & {Evidence shall show:
                \begin{enumerate}[label=(\alph*)]
                    \item that the most recent analysis was conducted less than one year prior to the start date of this audit, or after a major update to the tool,
                \end{enumerate}
                unless such update was more than one year prior to the start date of this audit, in which case, evidence shall show:
                \begin{enumerate}[resume,label=(\alph*)]
                    \item justification for why such analysis is still appropriate for this audit.
                \end{enumerate}} \\
            Q.B.4 & Evidence shall show that the time span of the dataset is within one year of the start date of the analysis. \\

            \midrule
            \textit{Q.C} & \textit{The demographic categories for which disparate impact can be analyzed using the dataset shall be defined.} \\[\baselineskip]
            Q.C.1 & Evidence shall identify the demographic categories for which disparate impact can be analyzed. \\
            Q.C.2 & Evidence shall show that such demographic categories include at the minimum: race/ethnicity and gender. \\
            Q.C.3 & Evidence shall disclose the method by which demographic data was collected. \\
            Q.C.4 & Evidence shall identify and disclose the demographic categories that are out of scope for this analysis. \\
            Q.C.5 & {Where demographic data were inferred, evidence shall:
                \begin{enumerate}[label=(\alph*)]
                    \item identify the method by which demographic data was inferred, and
                    \item show justification for why the selected method of demographic inference was appropriate.
                \end{enumerate}} \\
            
            \midrule
            \textit{Q.D} & \textit{Where the selection rate method is used, positive and negative outcomes of the tool shall be clearly defined as the basis for selection rate.} \\[\baselineskip]
            Q.D.1 & Evidence shall show justification for why this definition of positive outcome is appropriate. \\
            Q.D.2 & Where thresholding is used as a basis for positive outcome determination, evidence shall show justification for why the level(s) of threshold is (are) appropriate. \\
            Q.D.3 & {Evidence shall identify and disclose:
                \begin{enumerate}[label=(\alph*)]
                    \item all user-configurable tool settings,
                    \item whether each setting affects positive outcomes,
                \end{enumerate}
                and for all settings identified as outcome-affecting:
                \begin{enumerate}[resume,label=(\alph*)]
                    \item their extents of user configurability,
                    \item their default values, and
                    \item justification for why such default values were appropriate.
                \end{enumerate}} \\
            Q.D.4 & Evidence shall disclose the user-configurable tool settings and combinations of settings used for the analysis. \\

            \midrule
        
        \end{tabular}
    
    \end{table*}

    \begin{table*}[ht]
            
        \begin{tabular}{lp{0.8\textwidth}}

            \toprule
            ID & \textit{Criterion} \& Sub-criterion \\ 

            \midrule
            \textit{Q.E} & \textit{A metric which corresponds to selection rate or scoring rate shall be defined.} \\[\baselineskip]
            Q.E.1 & Where the selection rate method is used, evidence shall show that the selection rate of a group is defined as the ratio of positive outcome to all outcomes for that group. \\
            Q.E.2 & Where the scoring rate method is used, evidence shall show that the scoring rate of a group is defined as the rate at which that group receives a score from the tool above the median score of the sample. \\
        
            \midrule
            \textit{Q.F} & \textit{The ‘favored group’ and ‘disfavored groups’ shall be identified, for all demographic categories.\footnote{‘Favored group’ refers to the group with the highest selection or scoring rate in the analysis, and ‘disfavored groups’ refer to all other groups in the demographic category (i.e., with lower selection or scoring rate) \cite{ugesp}.}} \\[\baselineskip]
            Q.F.1 & Evidence shall show that the favored and disfavored groups are identified based on selection rates or scoring rates. \\
            Q.F.2 & Evidence shall show that the groups pertaining to race/ethnicity satisfy \S 60-3.4 B of the EEO guidelines \cite{disp}. \\
            Q.F.3 & {Where the groups pertaining to race/ethnicity do not satisfy EEO guidelines, evidence shall show:
                \begin{enumerate}[label=(\alph*)]
                    \item justification for why such EEO grouping is not used, and
                    \item the appropriateness of any substituted grouping.
                \end{enumerate}} \\
            Q.F.4 & Evidence shall show that the groups pertaining to gender contains at least Male and Female. \\
            Q.F.5 & Evidence shall show that intersectional groups contain all permutations of race/ethnicity and gender groups. \\
            Q.F.6 & Where race/ethnicity and gender groups are not known for a sample of candidates assessed by the tool, the evidence shall disclose its sample size. \\
        
            \midrule
            \textit{Q.G} & \textit{The impact ratios shall be disclosed for all disfavored groups, for all demographic categories.} \\[\baselineskip]
            Q.G.1 & Where an impact ratio for a disfavored group is below 0.8, evidence shall show justification for why the disfavored group is disadvantaged. \\
            Q.G.2 & Evidence shall show results of uncertainty analysis (e.g., standard error for the mean) or error propagation of impact ratios in the form of absolute errors or error bars. \\
            Q.G.3 & Where demographic data was inferred, evidence shall show that systematic errors due to demographic inference are properly propagated in impact ratio calculations. \\
            Q.G.4 & {Where a gender, race/ethnicity, or intersectional group was excluded from impact ratio calculations due to its size being below 2\% of the total sample size, evidence shall show:
                \begin{enumerate}[label=(\alph*)]
                    \item justification for its exclusion,
                    \item its sample size, and
                    \item its selection rate or scoring rate.
                \end{enumerate}} \\
        
            \midrule
            \textit{Q.H} & \textit{Where the selection rate method was used, statistical significance calculations of the difference between selection rates shall satisfy Uniform Guidelines on Employee Selection Procedures (UGESP) \cite{ugesp}.} \\[\baselineskip]
            Q.H.1 & Evidence shall show that statistical significance is calculated using the Two Independent-Sample Binomial Z-Test for sample sizes of 30 or more, and using the Fisher’s Exact Test for sample sizes of fewer than 30. \\

            \midrule
            \textbf{G} & \multicolumn{1}{c}{\textbf{Governance}} \\ 
            
            \midrule
            \textit{G.A} & \textit{The auditee shall have a party which is accountable for risks related to disparate impact.} \\[\baselineskip]
            G.A.1 & Evidence should show that the accountable party is a committee, but may also show that the accountable party is a single individual. \\
            G.A.2 & Evidence shall clearly show that risks related to disparate impact are owned and managed by the accountable party. \\

            \midrule
            
        \end{tabular}
    
    \end{table*}
            
    \begin{table*}[ht]
            
        \begin{tabular}{lp{0.8\textwidth}}

            \toprule
            ID & \textit{Criterion} \& Sub-criterion \\ 

            \midrule
            \textit{G.B} & \textit{The duties of the party accountable for risks related to disparate impact shall be clearly defined.} \\[\baselineskip]
            G.B.1 & Evidence shall show that such duties pertain to the ownership, management, and monitoring of risks related to disparate impact. \\
            G.B.2 & Evidence shall show that the accountable party has influence over product changes per effective challenge in Federal Guidance on Model Risk Management \cite{sr117}. \\
    
            \midrule
            \textit{G.C} & \textit{The auditee shall provide evidence that the defined duties of the party accountable for risks related to disparate impact are carried out.} \\[\baselineskip]
            G.C.1 & Evidence shall show that the defined duties were carried out prior to the start date of this audit. \\
            
            \midrule
            \textbf{R} & \multicolumn{1}{c}{\textbf{Risk Assessment}}\\ 
            
            \midrule
            \textit{R.A} & \textit{The auditee shall have completed a risk assessment of the tool.} \\[\baselineskip]
            R.A.1 & Evidence shall show that a risk assessment or an equivalent analysis was completed less than one year prior to the issuance date of this audit. \\
    
            \midrule
            \textit{R.B} & \textit{The risk assessment shall show identification of relevant risks related to bias.} \\[\baselineskip]
            R.B.1 & Evidence shall show the identification of risks related to various biases along all stages of the AI lifecycle, such as listed in the National Institute of Standards and Technology (NIST) Standard for Identifying and Managing Bias in Artificial Intelligence \cite{schwartz_22}. \\
            R.B.2 & Evidence shall show awareness of the parties potentially affected by the decisions made along all stages of the AI lifecycle. \\
    
            \midrule
            \textit{R.C} & \textit{The risk assessment shall demonstrate appropriate evaluation of relevant risks.} \\[\baselineskip]
            R.C.1 & Evidence shall show that the identified risks are assessed from the perspectives of multiple affected  external and internal stakeholders, with justifications for the extent of and mechanism by which such risks affect these stakeholders. \\
            R.C.2 & {Evidence shall show that the identified risks are assessed:
                \begin{enumerate}[label=(\alph*)]
                    \item in a sufficiently rigorous manner, using a quantitative and/or qualitative evaluation scheme, and
                    \item along multiple dimensions, such as but not limited to likelihood of harm and severity of harm.
                \end{enumerate}} \\
            R.C.3 & Evidence shall show justification for the provided evaluation of risks. \\
            
            \bottomrule

        \end{tabular}
    
    \end{table*}

\end{document}